\documentclass[journal]{IEEEtran}

\ifCLASSINFOpdf
\else
\fi
\usepackage{graphicx}
\usepackage{setspace}
\usepackage{amsmath,empheq}
\usepackage{amssymb}
\usepackage{amsthm}
\usepackage{caption}
\usepackage{subcaption}
\usepackage{bm} 
\usepackage{cite}
\usepackage{float}
\usepackage[usenames,dvipsnames]{pstricks}
\usepackage{epsfig}
\usepackage{pst-grad} 
\usepackage{pst-plot} 
\usepackage{tabularx}
\RequirePackage{flushend}
\usepackage{amsmath}
\usepackage[T1]{fontenc}

\newcommand{\nn} {\nonumber}
\usepackage{xcolor}
\usepackage{lipsum}
\usepackage{comment}

\allowdisplaybreaks

\allowdisplaybreaks
\usepackage[utf8]{inputenc}

 \IEEEaftertitletext{\vspace{-2.5\baselineskip}}
 \usepackage{svg}

\usepackage{algpseudocode}
\usepackage{algorithm}
\usepackage{xcolor}
\algnewcommand\algorithmicforeach{\textbf{for each}}
\algdef{S}[FOR]{ForEach}[1]{\algorithmicforeach\ #1\ \algorithmicdo} 

\begin{document}
\bstctlcite{IEEEexample:BSTcontrol}
\title{Joint Transmit and Jamming Power Optimization for Secrecy in Energy Harvesting Networks: A Reinforcement Learning Approach}
  
  \author{Shalini Tripathi,~\IEEEmembership{Student Member,~IEEE,} Chinmoy Kundu,~\IEEEmembership{Member,~IEEE,} Animesh Yadav,~\IEEEmembership{Senior Member,~IEEE,}  Ankur Bansal,~\IEEEmembership{Senior Member, IEEE}, Holger Claussen,~\IEEEmembership{Fellow,~IEEE,} and Lester Ho,~\IEEEmembership{Member,~IEEE}

\thanks{This work was supported in part by Taighde Éireann – Research Ireland under Grant number 22/PATH-S/10788.} 

\thanks{Shalini Tripathi and Ankur Bansal are with Indian Institute of Technology Jammu, Jammu 181221, India (e-mail:2019REE0001@iitjammu.ac.in; ankur.bansal@iitjammu.ac.in).}
\thanks{Chinmoy Kundu, Holger Claussen, and Lester Ho are with the Wireless Communications Laboratory, Tyndall National Institute, Dublin, Ireland. Holger Claussen is also with University College Cork, Cork, Ireland and Trinity College Dublin, Dublin, Ireland (e-mail: chinmoy.kundu@tyndall.ie, holger.claussen@tyndall.ie, lester.ho@tyndall.ie).}
\thanks{Animesh Yadav is with the School of Electrical Engineering and Computer Science, Ohio University, Athens, OH, USA (e-mail: yadava@ohio.edu).}
}



         
     

\maketitle
 \thispagestyle{empty}
\pagestyle{plain} 

\begin{abstract}
In this paper, we address the problem of joint allocation of transmit and jamming power at the source and destination, respectively, to enhance the long-term cumulative secrecy performance of an energy-harvesting wireless communication system until it stops functioning in the presence of an eavesdropper. 
The source and destination have energy-harvesting devices with limited battery capacities. The destination also has a full-duplex transceiver to transmit jamming signals for secrecy.  
We frame the problem as an infinite-horizon Markov decision process (MDP) problem and propose a reinforcement learning (RL)-based optimal joint power allocation (OJPA) algorithm that employs a policy iteration (PI) algorithm. 
Since the optimal algorithm is computationally expensive, we develop a low-complexity sub-optimal joint power allocation (SJPA) algorithm, namely, reduced state joint power allocation (RSJPA). Two other SJPA algorithms, the greedy algorithm (GA), and the naive algorithm (NA) are implemented as benchmarks.  In addition, the OJPA algorithm outperforms the individual power allocation (IPA) algorithms termed individual transmit power allocation (ITPA) and individual jamming power allocation (IJPA), where the transmit and jamming powers, respectively, are optimized individually. The results show that the OJPA algorithm is also more energy efficient.
Results also show that the OJPA algorithm significantly improves the secrecy performance compared to all SJPA algorithms. The OJPA algorithm also outperforms the secrecy performance of a genetic algorithm-based RL algorithm and a finite-horizon RL algorithm.
The proposed RSJPA algorithm achieves nearly optimal performance with significantly less computational complexity marking it the balanced choice between the complexity and the performance. 
We find that the computational time for the RSJPA algorithm with considering only $50$ percent of the total number of states is around $75$ percent less than the OJPA algorithm.

\end{abstract}
\begin{IEEEkeywords}
Energy harvesting, physical layer security, Markov decision process, reinforcement learning, policy iteration, full-duplex.
\end{IEEEkeywords}
\IEEEpeerreviewmaketitle
\vspace{-.2cm}

\section{Introduction}

Wireless sensor networks (WSNs) or Internet-of-Things (IoT) networks consist of numerous spatially distributed transmitting and receiving nodes designed to monitor physical phenomena or cooperatively exchange data between nodes. These networks are employed for various real-time applications 
such as multimedia surveillance, environmental monitoring, advanced healthcare delivery, industrial process control, smart homes, border surveillance, vehicle tracking, etc \cite{xu2015_WSN_intro, sah2022_WSN_intro}. Typically, these nodes rely on non-rechargeable batteries with constrained energy storage. However, many applications that employ these nodes, require continuous operation over an extended period without the possibility of replacing batteries. This poses a significant challenge for keeping networks operational for an extended period. Consequently, energy management strategies have been developed to manage limited energy resources carefully and extend the operational lifetime of these wireless networks \cite{xu2015_WSN_intro}.  

Recent advancements in hardware design have enabled the potential application of energy harvesting (EH) technology in wireless systems. EH in wireless systems can prolong the operating lifetime of networks 
\cite{sudevalayam2010EH_paper, ku2015EH_paper}. 
EH allows nodes to accumulate energy from ambient sources like solar, wind, and vibrations in their rechargeable batteries, unlike standard battery-powered transceivers with limited battery capacity and lifetime. 
However, the EH rate is often irregular, and EH devices are susceptible to physical destruction or hardware failure. 
Besides the aforementioned limitations, wireless channels are also time-varying in nature. 
Therefore, the problem at hand is to determine the strategy for transmit power allocation per time slot of EH transmitter in wireless networks in order to optimize the long-term cumulative performance of the network over its lifespan. This optimization must take into account intermittent energy arrivals at the EH nodes, the amount of energy available in the battery, and the prevailing channel conditions \cite{dohler2013learning}. 

To optimize the transmit power strategy of an EH transmitter in a wireless network, conventional optimization approaches solely focus on single time slot optimization problems or greedy approaches to maximize the immediate reward \cite{azarhava2020_conv_optimization, salari2023_conv_optimization}. 
The article \cite{azarhava2020_conv_optimization} 
considers a wireless energy harvesting sensor network consisting of a hybrid access point (HAP) with an unlimited power supply and multiple EH sensors. These sensors harvest energy in the downlink from the HAP and then transmit sensed data to the HAP in the uplink. 
The energy efficiency of the network is maximized by optimizing the duration of EH, the duration of the data transmission of sensors, and the transmit power allocation for the sensors. In \cite{salari2023_conv_optimization}, an EH cooperative cognitive radio network consisting of two transceivers and 
multiple two-way amplify-and-forward (AF)-based relays is considered. 
The relays periodically switch between EH and the information transmission phase. 
The work aims to maximize the secondary network's sum rate by jointly obtaining the optimum EH time allocation and the distributed beamforming vector for the relays.
These conventional optimization approaches can not optimize the long-term cumulative performance until the network is no longer operational due to lack of energy  \cite{Octavia2019}. 
To maximize the cumulative performance, often called cumulative reward, a causal problem needs to be formulated where only the past and current knowledge of the system state are available, with no foresight into the future \cite{yadav2017_TCOM_intro, Ajib_VTC}. Such causal problems are, essentially, a sequential decision making problem that make decisions without the knowledge of the future, and can be solved optimally via dynamic programming and reinforcement learning (RL) techniques \cite{sutton1998RL, puterman2014markov}. RL is a machine learning technique where an agent learns to make decisions by interacting with its environment and receiving feedback in the form of rewards. By exploring different actions and observing the reward values, the agent gradually learns to take actions that maximize the cumulative reward \cite{sutton1998RL}. RL has the advantage of providing the optimal solution without the knowledge of future information \cite{sutton1998RL, puterman2014markov}.

To maximize the long-term cumulative performance in EH wireless networks, the application of RL is considered in \cite{wong2012ICC,kashef2012optimal,dohler2013learning,wong2014joint}. In \cite{wong2012ICC}, a wireless sensor network is considered where the sensor node is equipped with EH capability and a finite data buffer. Optimal energy allocation for sensing and transmission for the sensor node is obtained by maximizing the total throughput over a finite horizon of time. The solution is achieved by formulating a finite-horizon Markov decision process (MDP) problem using the backward induction algorithm. In \cite{kashef2012optimal}, a point-to-point wireless system with EH source equipped with an infinite energy queue is considered. An MDP problem is formulated to decide whether the EH source transmit or not in a given time slot. 
The objective is to maximize the average number of successfully delivered packets per time slot by the source. 
In \cite{dohler2013learning}, an EH transmitter with a finite battery capacity is considered. The MDP-based problem is formulated that maximizes the expected total transmitted data over the lifetime of the transmitter under the finite battery capacity constraint.
The system in \cite{wong2014joint} considers a transmitter with a finite data buffer and energy consumption in data sensing in addition to the system features assumed in \cite{kashef2012optimal, dohler2013learning}. 
A joint energy allocation strategy for transmission and sensing is obtained to maximize the expected total amount of data transmitted until the transmitter stops functioning.
Problems formulated in \cite{kashef2012optimal,dohler2013learning,wong2014joint} are infinite-horizon MDP, whereas the solutions are provided using value iteration in \cite{kashef2012optimal,wong2014joint} or policy iteration (PI) in \cite{dohler2013learning}. It is to be noted that the PI-based algorithms converge faster than the value iteration-based algorithms \cite{PI_heydari2016}.

Aforementioned works assume no attack on wireless signal that compromises data privacy. However, wireless signals are susceptible to attacks by unauthorized users, e.g., transmission interception by an eavesdropper and disruption by a jammer \cite{zou2016survey}.
An active eavesdropper may employ full-duplex mode for jamming to degrade legitimate reception and eavesdropping simultaneously \cite{zhou2012PLS}. Where nodes have limited energy and computational capabilities, such as in an IoT use case, employing security using physical layer security (PLS) techniques can be a viable approach to enhance security due to its low complexity \cite{shiu2011PLS}. PLS does not require secret keys, and thus, eliminating the complexities associated with key generation, distribution, and management associated with cryptographic security. PLS exploits the inherent randomness and imperfections present in the wireless channel to provide security \cite{wyner_wiretap}. 

To address the problem of secrecy through PLS, recently, a few works \cite{hoseini2023GC, yang2020_PLS_drl_TWC, saleem2022_PLS_drl, liu2024secrecy_DRL, 
li2022DRL, 
yang2024EH_DRL, qian2022secrecy} have employed RL framework for solution. A large wireless network with multiple access points (APs), users, and eavesdroppers is considered in \cite{hoseini2023GC}. If an AP has no associated user to receive data, it works as a jammer.  The joint optimal user association and power allocation to the APs are considered by maximizing the sum secrecy capacity of the users. The soft actor-critic algorithm from the RL framework is proposed as a solution. A reconfigurable intelligent surface (RIS)-aided wireless system with a base station (BS) and multiple users in the presence of
multiple eavesdroppers is considered in \cite{yang2020_PLS_drl_TWC}. To improve the secrecy rate of the system, a design problem is formulated to jointly optimize the BS’s beamforming and the RIS’s reflecting beamforming. A deep reinforcement
learning (DRL)-based technique is proposed to achieve a secure beamforming policy in dynamic time-varying channel conditions.
A multiple-input single-output (MISO) downlink system is considered in \cite{saleem2022_PLS_drl} where a BS with multiple transmit antennas communicates with multiple single-antenna devices with the help of an RIS. Legitimate devices are classified into trusted and untrusted devices, where the untrusted devices may potentially eavesdrop on the trusted devices. 
A deep deterministic policy gradient (DDPG)-based  RL algorithm is proposed to obtain the joint optimal RIS phases and transmit beamforming by maximizing the sum secrecy rate of trusted devices while ensuring performance guarantee to all trusted and untrusted devices. 

In \cite{liu2024secrecy_DRL}, a secure Visible Light Communication (VLC) system is considered, where multiple light fixtures serve as friendly jammers. A DRL algorithm is implemented to optimize the friendly jamming policy, assuming continuous state and action spaces.
In \cite{li2022DRL}, a smart cyber-attack scenario is examined, where attackers can dynamically select their attack methods, such as jamming or eavesdropping. An RL solution is employed to predict the attack strategies and intelligently determine whether artificial noise should be added to the transmitted signal.
A Reconfigurable Intelligent Surface (RIS)-mounted Unmanned Aerial Vehicle (UAV)-assisted maritime communication system under jamming attacks is analyzed in \cite{yang2024EH_DRL}. To jointly optimize the transmission power of the base station, the placement of the UAV-RIS, and the RIS's reflecting beamforming, a DRL-based approach is proposed.
In \cite{qian2022secrecy}, a UAV-aided Non-Orthogonal Multiple Access (NOMA) system is investigated for data collection from Transmission Devices (TDs) in the presence of an eavesdropping attack. A group of Auxiliary Devices (ADs) is deployed to provide cooperative jamming against the eavesdropper. A DRL-based online optimization algorithm is introduced to maximize the total secrecy capacity by jointly optimizing the power allocations of TDs and ADs.

None of the aforementioned works that address secrecy of networks using RL approaches in \cite{hoseini2023GC, yang2020_PLS_drl_TWC, saleem2022_PLS_drl, liu2024secrecy_DRL, 
li2022DRL, 
yang2024EH_DRL, qian2022secrecy} consider the EH capability at the source or the destination nodes, which is essential for extending the lifespan of a network. To determine the potential impact of current decisions on the future secrecy performance due to energy limitation in EH networks, a communication system consisting of an EH source node, a full-duplex destination node, and a full-duplex active eavesdropper is considered in \cite{insoo2019FD}. A self-interference attenuation factor is considered at the full-duplex nodes reflecting the difficulty of fully suppressing own transmit signal. The destination node transmits an artificial noise while decoding the transmitted signal from the source only if the eavesdropper does not transmit a jamming signal.  An optimal source transmit power decision policy is obtained to maximize the long-term secrecy rate using the value iteration algorithm in the RL framework. 

In the systems with EH nodes, where cumulative performance is optimized, secrecy was not a concern \cite{kashef2012optimal,dohler2013learning,wong2014joint,wong2012ICC}. Though \cite{insoo2019FD} achieves secrecy through full-duplex destination jamming, the article only focuses on optimizing the transmit power of the source. In EH wireless networks where both the source and the destination rely solely on energy harvesting and the full-duplex destination is jamming for secrecy,
there is a necessity to optimize jamming power jointly with the source transmit power to enhance the long-term cumulative secrecy performance.
If too little jamming power is assigned, the jammer cannot launch an effective jamming attack, conversely, excessive power to create jamming attacks can exhaust the jammer's battery and increase self-interference. Both may lead to a decreased long-term cumulative secrecy performance. One should find the right balance by assigning the optimal amount of power to both the transmitter and the jammer jointly.

Motivated by the above discussion, we consider a wireless communication system consisting of a EH source, an EH destination, and an eavesdropper. The destination has the full-duplex capability to simultaneously receive and produce a jamming attack to disrupt eavesdropping. The network operates in discrete time slots.
We assume that the wireless channel between any nodes remains constant within a time slot; however vary between consecutive time slots following the first-order discrete-time  Markov model. The arrival of energy packets is modeled as a Bernoulli process. For the system to be more realistic, we assume that the network might stop being operational due to physical destruction and hardware failure at any time slot with a certain probability. As a result, the lifetime of the network becomes a random variable. For the considered system, we maximize the long-term expected total transmitted secure bits until the network stops functioning by jointly allocating power for source transmission and destination jamming. 
The proposed joint power allocation takes into account the probability that the network remains operational at each time slot, battery energy level, EH rate, channel conditions, and self-interference attenuation factor at the destination. 

The main contributions of the paper are outlined as follows: 

\begin{itemize}

\item We study, for the first time, the optimal joint power allocation (OJPA) problem for source transmission and destination jamming to maximize the long-term expected total transmitted secure bits where both the source and destination are energy harvesting in an EH wireless network until the network stops functioning using the RL framework.  The problem is formulated as an infinite-horizon MDP as the lifetime of the proposed network is a random variable. The proposed OJPA algorithm utilizes the PI algorithm for the solution due to its faster convergence.
       
\item We propose a low computational complexity sub-optimal joint power allocation (SJPA) algorithm, namely, the reduced state joint power allocation (RSJPA), which is partially based on the PI algorithm with a smaller subset of the system states. Two other SJPA algorithms, greedy algorithm (GA) and naive algorithm (NA), are also implemented. Besides, we also develope two individual power allocation (IPA) algorithms (i.e., individual transmit power allocation (ITPA) and individual jamming power allocation (IJPA)) designed using the same RL framework). In the ITPA algorithm, the transmit power is optimized with a fixed destination power supply, and in the IJPA algorithm, the jamming power is optimized with a fixed source power supply.  

\item 
Additionally, we compare the secrecy performance of the proposed RL algorithms (OJPA and RSJPA) with that of a genetic algorithm-based RL algorithm. The performance of the OJPA algorithm (infinite-horizon) is also compared with that of a finite-horizon RL algorithm. The results show that the OJPA algorithm outperforms both the genetic algorithm-based and finite-horizon RL algorithms.

\color{black}

\item We derive the computational complexity of the OJPA and SJPA algorithms, and present a comprehensive performance comparison of the SJPA and IPA algorithms. It is found that the proposed OJPA algorithm not only maximizes the long-term expected total transmitted secure bits but is also most energy-efficient.

\end{itemize}

The rest of the paper is organized as follows: Section II describes the system model. Section III formulates the problem of joint transmit and jamming power allocation for the source transmission and destination nodes, respectively. Section IV proposes RL-based OJPA and SJPA solution approaches. Section V describes the two IPA algorithms, and Sections VI and VII provide computational complexity and numerical results, respectively. Finally, Section VIII concludes the paper. 

\textit{Notation:}  
$\mathbb{P[\cdot]}$ denotes the probability of an event, $\mathbb{E}\left[\cdot \right]$ denotes the expectation operator.
$\max\{\cdot\}$ and $\min\{\cdot\}$ denote the maximum and minimum of its arguments, respectively.


\section{System Model}


 \begin{figure}
 \raggedleft
  \includegraphics[width=3in]{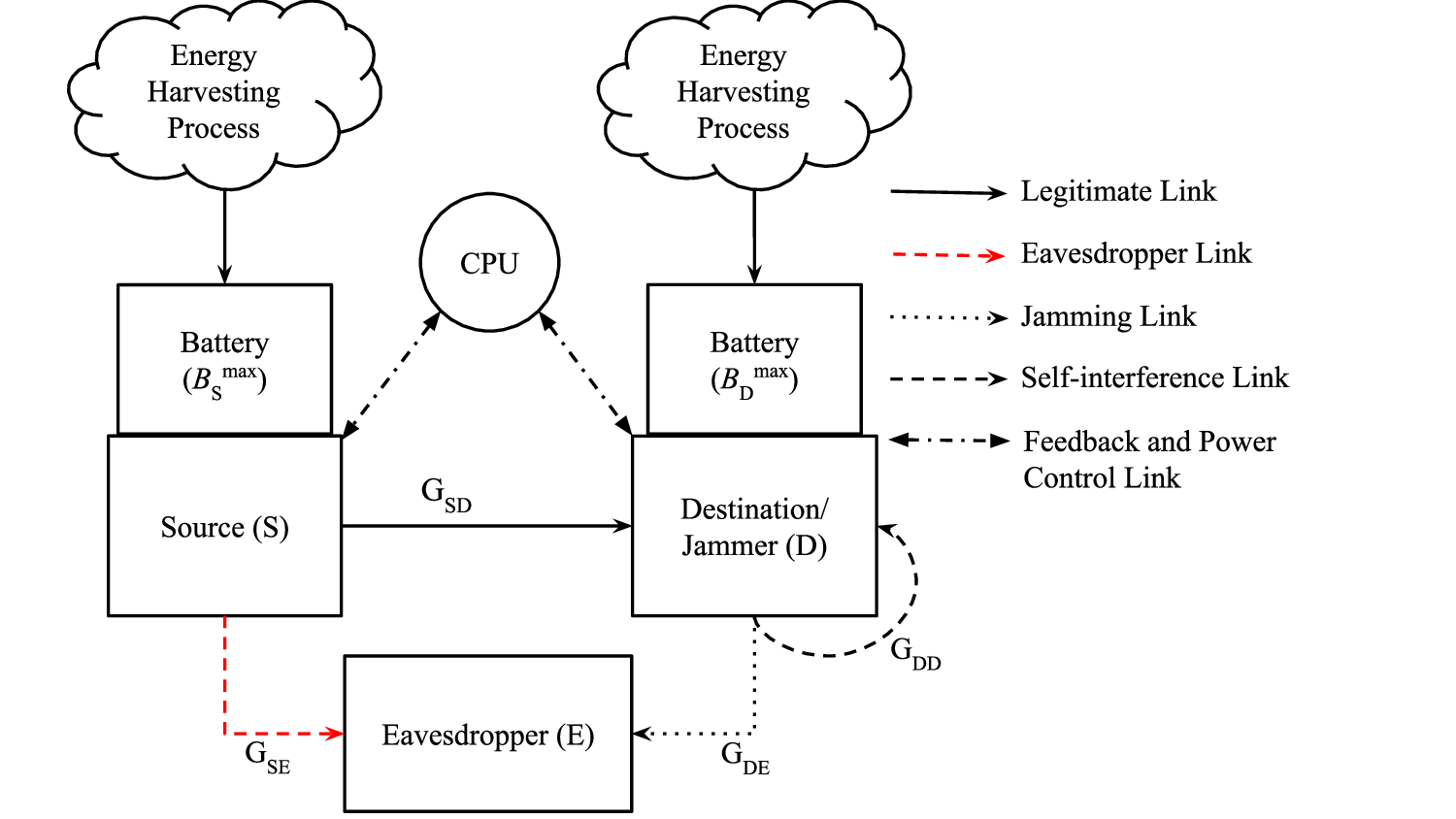}
 \caption{A wireless system with an EH source, an EH full-duplex destination, and a passive eavesdropper.}
 \label{fig_system}
 \vspace{-0.4cm}
\end{figure}

Consider an EH wireless communication system where a source node $\textrm{S}$ is communicating with a destination node $\textrm{D}$ in the presence of a passive eavesdropping node $\textrm{E}$, as illustrated in Fig. \ref{fig_system}. 
We assume that both nodes $\textrm{S}$ and $\textrm{D}$ are equipped with an EH device that contains a rechargeable battery with limited storage capacity, whereas node $\textrm{E}$ is equipped with a regular power supply from the traditional power grid. Furthermore, node $\textrm{D}$ operates in a full-duplex mode, and nodes S and E operate in a half-duplex mode. The full-duplex mode enables node D to receive data from node $\textrm{S}$ and simultaneously perform jamming attacks on node $\textrm{E}$. Simultaneous jamming attack causes self-interference to signal reception at node $\textrm{D}$. We assume that a mechanism for self-interference cancellation (SIC) is in-place at node D; however, residual self-interference remains and is captured through a factor $0\leq \alpha \leq 1$, where $\alpha=1$ implies no SIC is performed and $\alpha=0$ indicates SI is canceled completely \cite{insoo2019FD}. We assume that a central processing unit (CPU) exists in the network which has access to global channel state and battery state information, and performs power allocation decisions with the help of this information. 

Transmission occurs in a time-slotted manner over the period of $K$ time slots (TSs), and each TS is indexed by $k \in \mathcal{K} = \{0,1,\ldots, K-1\}$.
The TSs have an identical duration of $T_s$ seconds \cite{wong2012ICC, wong2014joint, dohler2013learning, Octavia2019}. 
We assume that \textrm{S} has a sufficient amount of data available for transmission in each TS. Since EH nodes are susceptible to physical destruction or hardware failure, we assume the network lifetime $K$ is a random variable.
Let $\Gamma \in [0,1)$ be the probability that the network remains operational throughout a given TS 
enduring physical damage or hardware malfunctions, where $\Gamma$ is constant for each of TS. Accordingly, the network lifetime $K$ can be modeled as a geometrically distributed random variable with mean $1/(1-\Gamma)$ \cite{wong2014joint, dohler2013learning}. 

\subsection{EH Model} 
We assume that the energy harvested by node $\textrm{S}$ and node $\textrm{D}$ in the $k$th TS are $H_\textrm{S}^{(k)}\in \mathcal{H}_\textrm{S}$ and $H_\textrm{D}^{(k)}\in \mathcal{H}_\textrm{D}$ energy units, respectively, where $\mathcal{H}_\textrm{S} = \{0,E_\textrm{S}\}$ and $\mathcal{H}_\textrm{D} = \{0,E_\textrm{D}\}$ is the set of possible harvested energy units.
We model the EH event in each TS at node $\textrm{S}$ and node $\textrm{D}$ as an independent and identically distributed Bernoulli process with probability $p$ and $q$, respectively, independent of data transmission process   \cite{octavia2020VTC}. 
Accordingly, in each TS $k\in\mathcal{K}$, the probability of harvesting energy of $E_\textrm{S}$ and $E_\textrm{D}$ energy units at node $\textrm{S}$ and node $\textrm{D}$, respectively, is $\mathbb{P}[H_\textrm{S}^{(k)}=E_\textrm{S}]=p$ and $\mathbb{P}[H_\textrm{D}^{(k)}=E_\textrm{D}]=q$, respectively,  and the probability of not harvesting any energy is $\mathbb{P}[H_\textrm{S}^{(k)}=0]=1-p$ and $\mathbb{P}[H_\textrm{D}^{(k)}=0]=1-q$, respectively.

The battery capacities of node $\textrm{S}$ and node $\textrm{D}$ are $B_{\textrm{S}}^{\textrm{max}}$ and $B_{\textrm{D}}^{\textrm{max}}$ energy units, respectively.
The amount of energy stored in the battery of node $\textrm{S}$ and node $\textrm{D}$ in the $k$th TS is $B_\textrm{S}^{(k)}\in \mathcal{B}_\textrm{S}$ and $B_\textrm{D}^{(k)}\in \mathcal{B}_\textrm{D}$ energy units, respectively, where  $\mathcal{B}_\textrm{S} = \{0,1, \ldots, B_{\textrm{S}}^{\textrm{max}}\}$ and $\mathcal{B}_\textrm{D} = \{0,1, \ldots, B_{\textrm{D}}^{\textrm{max}}\}$ are the set of possible discrete energy levels.

Note that the energy utilized for the signal transmission or the jamming attack during the $k$th TS can not exceed the amount of energy stored in corresponding batteries. Similarly, the storage of harvested energy is limited by the battery capacity. As a result, the energy levels $B_\textrm{S}^{(k+1)}$ and $B_\textrm{D}^{(k+1)}$ in the $(k+1)$-th TS is updated from the $k$th TS as

\begin{align}
\label{battery_Bs}
   B_\textrm{S}^{(k+1)} 
   \hspace{-0.1cm}
   &= 
   \hspace{-0.1cm}
   \left\{ 
    \hspace{-0.2cm}
   \begin{array}{lcl}
    \min
    \{ B_\textrm{S}^{(k)} - P_\textrm{S}^{(k)}  T_s + E_\textrm{S},B_{\textrm{S}}^{\textrm{max}} \}
    \,\, \mbox{for}\,\,H_\textrm{S}^{(k)} \hspace{-0.1cm}=\hspace{-0.1cm} E_\textrm{S}  
    \\ B_\textrm{S}^{(k)} - P_\textrm{S}^{(k)} T_s 
   \hspace{2.9cm} \mbox{for}\,\,H_\textrm{S}^{(k)} = 0 
    \end{array}
    \right.  \\
\label{battery_Bd}
   B_\textrm{D}^{(k+1)} 
   \hspace{-0.1cm}
   &= 
   \hspace{-0.1cm}
   \left\{ 
    \hspace{-0.2cm}
   \begin{array}{lcl}
    \min
    \{ B_\textrm{D}^{(k)} - P_\textrm{D}^{(k)}  T_s + E_\textrm{D},B_{\textrm{D}}^{\textrm{max}} \}
    \,\, \mbox{for}\,\,H_\textrm{D}^{(k)} \hspace{-0.1cm}=\hspace{-0.1cm} E_\textrm{D} 
    \\ B_\textrm{D}^{(k)} - P_\textrm{D}^{(k)} T_s 
   \hspace{2.9cm} \mbox{for}\,\,H_\textrm{D}^{(k)} = 0 
    \end{array}
    \right. 
\end{align}
where $P_\textrm{S}^{(k)} \in \mathcal{P}$ and $ P_\textrm{D}^{(k)} \in \mathcal{P}$ are the power transmitted by node $\textrm{S}$ and node $\textrm{D}$ for signal transmission and jamming attack, respectively, in the $k$th TS, $\mathcal{P} = \{P_1,P_2, \ldots, P_M\}$ is the set of $M$ possible transmit power levels, and $P_\textrm{S}^{(k)} T_s\leq B_\textrm{S}^{(k)}$, $P_\textrm{D}^{(k)} T_s\leq B_\textrm{D}^{(k)}$. Here we note that the EH process and data transmission are occurring simultaneously. The harvested energy $H_\textrm{S}^{(k)}$ in the $k$th TS will be available to use in the $(k+1)$th or later TSs.

\subsection{Channel and Signal Transmission Model}

We denote the channel power gain of a link XY in the $k$th TS by $G_{\textrm{XY}}^{(k)}$, where $\textrm{XY} \in\{\textrm{SD, SE, DD, DE}\}$ is the link between any possible nodes $\textrm{X}\in\{\textrm{S, D}\}$ and $\textrm{Y}\in\{\textrm{D, E}\}$. The self-interference link at node $\textrm{D}$ is denoted as $\textrm{DD}$. We consider that $G_{\textrm{XY}}^{(k)}$ for any $k$th TS is quantized to $L$ finite levels, i.e, {$G_{\textrm{XY}}^{(k)}\in\mathcal{G}$, where $\mathcal{G}=\{G_1, G_2, \ldots, G_L\}$ is a set of $L$ discrete values.
The channel power gain remains unchanged during a particular TS, however, it transitions to a new value in the next TS taking values from $\mathcal{G}$. We assume that the channel state transition follows a first-order Markov model \cite{dohler2013learning}. The Markov model incorporates the uncertainty of the wireless propagation environment.

The signals received at node $\textrm{D}$ and node $\textrm{E}$ in the $k$th TS slot can be expressed, respectively, as
\begin{align}
\label{EQ_receved_signal}
    y_\textrm{D}^{(k)} &= \sqrt{G_{\textrm{SD}}^{(k)} P_\textrm{S}^{(k)}} x_\textrm{S}^{(k)} +  \sqrt{\alpha G_{\textrm{DD}}^{(k)} P_\textrm{D}^{(k)}} w_\textrm{D}^{(k)} + z_\textrm{D}^{(k)},\\    
    y_\textrm{E}^{(k)} &= \sqrt{G_{\textrm{SE}}^{(k)} P_\textrm{S}^{(k)}} x_\textrm{S}^{(k)} + \sqrt{G_{\textrm{DE}}^{(k)} P_\textrm{D}^{(k)}} w_\textrm{D}^{(k)} + z_\textrm{E}^{(k)},
\end{align}
where $x_\textrm{S}^{(k)}$ and $w_\textrm{D}^{(k)}$ are the unit energy information signal and the jamming signal transmitted by node $\textrm{S}$ and node $\textrm{D}$, respectively. $P_\textrm{S}^{(k)}$ and $P_\textrm{D}^{(k)}$ are transmitted and jamming power used by node $\textrm{S}$ and node $\textrm{D}$, respectively. $\alpha$ is the self-interference attenuation factor, 
and $z_\textrm{D}^{(k)}$ and $z_\textrm{E}^{(k)}$ are the additive white Gaussian noise (AWGN) at node $\textrm{D}$ and node $\textrm{E}$ with zero mean and noise spectral density $N_0$ W/Hz. The corresponding signal-to-interference-plus-noise ratio (SINR) at node $\textrm{D}$ and node $\textrm{E}$ in the $k$th TS is expressed as
\begin{align}
\gamma_\textrm{D}^{(k)} =\frac{G_{\textrm{SD}}^{(k)} P_\textrm{S}^{(k)}}{\alpha P_\textrm{D}^{(k)} G_{\textrm{DD}}^{(k)} + W  N_0},
 ~~~\gamma_\textrm{E}^{(k)} =\frac{G_{\textrm{SE}}^{(k)} P_\textrm{S}^{(k)}}{P_\textrm{D}^{(k)} G_{\textrm{DE}}^{(k)} + W N_0},   
\end{align}
respectively, where $W$ is the bandwidth of the channel.

\subsection{Performance Metrics}
We now define the achievable secrecy rate in bits per second (bps) of the network in the $k$th TS as the difference in achievable rates between the destination channel and the eavesdropping channel as
\begin{align}
    C_\textrm{S}^{(k)} = \max \{C_\textrm{D}^{(k)} - C_\textrm{E}^{(k)},0\}\quad \text{bps},
    \label{eqCSk}
\end{align}
where $ C_\textrm{D}^{(k)} = W \log_2{(1+\gamma_\textrm{D}^{(k)})}$ and $C_\textrm{E}^{(k)} = W \log_2{(1+\gamma_\textrm{E}^{(k)})}$ are the achievable rates for the destination channel and the eavesdropping channel in the $k$th TS, respectively. The operator $\max\{\cdot\}$ in (\ref{eqCSk}) is to signify that the secrecy rate is always positive.
The expected total transmitted secure bits until the network stops functioning is defined as \cite{wong2014joint}

\begin{align}
    \mu =  \mathbb{E} \left[  \mathbb{E}_K \left[ \sum_{k=0}^{K-1}   C_\textrm{S}^{(k)} T_s \right] \right]\quad \text{bits},  
    \label{eqECS}
\end{align} 

where $\mathbb{E}_{K}[\cdot]$ denotes the expectation with respect to the random variable $K$ and $\mathbb{E}[\cdot]$ denotes the expectation taken over all other relevant random variables, i.e., $G_{\textrm{XY}}^{(k)}$ for all $\textrm{XY} \in\{\textrm{SD, SE, DD, DE}\}$, $H_{\textrm{S}}^{(k)}$, and $H_{\textrm{D}}^{(k)}$. 


\section{Problem Formulation}
The objective of the considered EH wireless communication system is to maximize the expected total transmitted secure bits $\mu$ in (\ref{eqECS}) by optimally allocating $P_\textrm{S}^{(k)}$ and $P_\textrm{D}^{(k)}$ in each TS until the network stops functioning.
The solution should consider the probability that the network remains operational at each TS, current battery energy level, EH rate, channel condition, and self-interference attenuation factor.
Accordingly, we formulate the problem of finding the joint power allocation for transmitting and jamming power as

\begin{subequations}
\begin{align}
    \text{P1}:  \underset{\{P^{(k)}_\textrm{S},P^{(k)}_\textrm{D}\}_{k = 0}^K }{\text{maximize}} 
     &  \mathbb{E} \left[  \mathbb{E}_K \left[ \sum_{k=0}^{K-1}   C_\textrm{S}^{(k)} T_s \right] \right]\label{joint_optimization_a}\\
     \text{s.t.}\,
     & (\ref{battery_Bs}), (\ref{battery_Bd})\label{joint_optimization_b}\\
     & 0  \leq P_\textrm{S}^{(k)} \leq \frac{B_{\textrm{S}}^{(k)}}{ T_s} \label{joint_optimization_c}\\
     & 0  \leq P_\textrm{D}^{(k)} \leq \frac{B_{\textrm{D}}^{(k)}}{T_s}. \label{joint_optimization_d}
\end{align}
\label{Problem}
\end{subequations}

The transmit and jamming power constraints are expressed in (\ref{joint_optimization_c}) and (\ref{joint_optimization_d}), respectively. A careful observation of the problem P1 reveals that the joint optimal allocation of powers at node $\textrm{S}$ and node $\textrm{D}$ not only depends on the knowledge of channel conditions and battery levels
in the current $k$th TS, it also depends on their values in the future TSs as well. 
As our system follows the Markov property, the formulated problem (\ref{Problem}) is an online sequential decision-making problem with finite action and state spaces with a bounded and consistent immediate reward function. Thus, we use an MDP-based framework to obtain a solution that aims to make optimal decisions at each decision epoch to maximize the expected total reward \cite{puterman2014markov} \cite{bellman1957markovian}.

\section{Proposed Solution}
\label{section_prposed_solution}
In this section, we develop optimal and sub-optimal solution strategies for (\ref{Problem}) using the MDP framework. The optimal solution strategy is discussed first, then computationally efficient sub-optimal strategies are described. 

\subsection{Preliminaries}
To understand the MDP-based solution approach, we first define five important terms related to an MDP framework, i.e., decision epochs, states, actions, state transition probabilities, and rewards (including immediate and expected discounted sum reward), in the context of problem (\ref{Problem}).  

\begin{itemize}

\item Decision epochs: The decision epochs are the TSs $k \in \mathcal{K}$ during which decisions are made. 

\item States: The states represent the collection of relevant information that describes the system under consideration. For our system, the state in TS 
$k$ is defined as $s^{(k)}=(G_{\textrm{SD}}^{(k)}, G_{\textrm{SE}}^{(k)}, G_{\textrm{DD}}^{(k)}, G_{\textrm{DE}}^{(k)}, \allowbreak B_\textrm{S}^{(k)}, B_\textrm{D}^{(k)})$. The state space is given by  $\mathcal{S} = \mathcal{G}_{\textrm{SD}} \times \mathcal{G}_{\textrm{SE}} \times \mathcal{G}_{\textrm{DD}} \times \mathcal{G}_{\textrm{DE}} \times \mathcal{B}_\textrm{S} \times \mathcal{B}_\textrm{D}$ with finite number of discrete possible states $N_S$. Here, $N_S=|\mathcal{S}|$, and $|\mathcal{S}|$ is the cardinality of the set $\mathcal{S}$. 

\item Actions: Actions are the collection of decisions available for the system that can be taken in TS $k$ for a given state $s^{(k)}$. For example, an action $a^{(k)}$ is taken to optimize the problem P1, i.e., a pair of transmit powers $\{P_\textrm{S}^{(k)},P_\textrm{D}^{(k)}\}$ is to be decided from the feasible action set $U(s^{(k)})$ such that 

\begin{align}
\label{eq_feasible_action}
      &a^{(k)} \in U(s^{(k)}) \nn \\
      &= \left\{P_\textrm{S}^{(k)},P_\textrm{D}^{(k)} \mid 0 \leq P_\textrm{S}^{(k)} \leq \frac{B_{\textrm{S}}^{(k)}}{T_s}, 0 \leq P_\textrm{D}^{(k)} \leq \frac{B_{\textrm{D}}^{(k)}}{T_s}\right\}. 
\end{align}

The action $a^{(k)}$ belongs to the set $\mathcal{A} =\{a_1,\ldots, a_{ N_A }\}$ of all possible actions where an action $a_i$ for any $i\in\{1,\ldots, N_A\}$ is the pair of transmit power levels $\{ P _m \in \mathcal{P},  P _n\in \mathcal{P}\}$ for any $m,n\in\{1,\ldots, M\}$, and $N_A =  M^2$ is the total number of possible actions. 
\item State transition probability: The state transition probability represents the probability of transitioning to the state $s^{(k+1)}$ from the state $s^{(k)}$ by taking an action $a^{(k)}$ in the $k$th TS which is expressed as 
\begin{align}
    &\mathbb{P}[s^{(k+1)}\mid s^{(k)}, a^{(k)}] \nn\\ 
    & = \mathbb{P}[G_{\textrm{SD}}^{(k+1)}, G_{\textrm{SE}}^{(k+1)}, G_{\textrm{DD}}^{(k+1)}, G_{\textrm{DE}}^{(k+1)}, 
 B_\textrm{S}^{(k+1)}, B_\textrm{D}^{(k+1)}\mid \nn\\ 
    &~~~~G_{\textrm{SD}}^{(k)}, G_{\textrm{SE}}^{(k)}, G_{\textrm{DD}}^{(k)}, G_{\textrm{DE}}^{(k)}, B_\textrm{S}^{(k)}, B_\textrm{D}^{(k)}, P_\textrm{S}^{(k)}, P_\textrm{D}^{(k)}] \nn\\ 
    &= \mathbb{P}[G_{\textrm{SD}}^{(k+1)}\mid G_{\textrm{SD}}^{(k)}] \times \mathbb{P}[G_{\textrm{SE}}^{(k+1)}\mid G_{\textrm{SE}}^{(k)}] \nn \\
    &~~~~\times \mathbb{P}[G_{\textrm{DD}}^{(k+1)}\mid G_{\textrm{DD}}^{(k)}] 
    \times \mathbb{P}[G_{\textrm{DE}}^{(k+1)}\mid G_{\textrm{DE}}^{(k)}] \nn \\
    &~~~~\times \mathbb{P}[B_\textrm{S}^{(k+1)} \mid B_\textrm{S}^{(k)}, H_\textrm{S}^{(k)}, P_\textrm{S}^{(k)}]
    \times\mathbb{P}[H_\textrm{S}^{(k)}]  \nn\\
    &~~~~\times \mathbb{P}[B_\textrm{D}^{(k+1)} \mid B_\textrm{D}^{(k)},  H_\textrm{D}^{(k)} , P_\textrm{D}^{(k)}]\times\mathbb{P}[H_\textrm{D}^{(k)}],
    \label{Transition_Prob}
\end{align}     
where $\mathbb{P}[B_\textrm{S}^{(k+1)} \mid B_\textrm{S}^{(k)}, H_\textrm{S}^{(k)}, P_\textrm{S}^{(k)}]$ and $\mathbb{P}[B_\textrm{D}^{(k+1)} \mid B_\textrm{D}^{(k)}, H_\textrm{D}^{(k)}, P_\textrm{D}^{(k)}]$ are equal to 1 if (\ref{battery_Bs}) and (\ref{battery_Bd}) are satisfied, zero otherwise. We also have $\mathbb{P}[H_\textrm{S}^{(k)}]=p$ when  $H_\textrm{S}^{(k)}=E_S$, $\mathbb{P}[H_\textrm{S}^{(k)}]=1-p$ when  $H_\textrm{S}^{(k)}=0$, $\mathbb{P}[H_\textrm{D}^{(k)}]=q$ when  $H_\textrm{D}^{(k)}=E_D$, and $\mathbb{P}[H_\textrm{D}^{(k)}]=1-q$ when  $H_\textrm{D}^{(k)}=0$.
If $\mathbb{P}[B_\textrm{S}^{(k+1)} \mid B_\textrm {S}^{(k)}, H_\textrm{S}^{(k)}, P_\textrm{S}^{(k)}]$ and $\mathbb{P}[B_\textrm{D}^{(k+1)} \mid B_\textrm{D}^{(k)}, H_\textrm{D}^{(k)}, P_\textrm{D}^{(k)}]$ are equal to zero, (\ref{Transition_Prob}) also becomes zero indicating the impossibility of a transition from state $s^{(k)}$ to state $s^{(k+1)}$ while taking action $a^{(k)}$. 

\item Rewards: When an action $a^{(k)}$ prompts a transition from $s^{(k)}$ to $s^{(k+1)}$ in the $k$th TS, it also results in an immediate reward $R^{(k)}(s^{(k)},a^{(k)})$. In the context of our problem, the immediate reward function in the $k$th TS from (\ref{eqCSk}) is 

\begin{align}
    R^{(k)}(s^{(k)},a^{(k)}) = C_\textrm{S}^{(k)} T_s,
    \label{reward}
\end{align}

and the expected total reward is expressed in (\ref{eqECS}). 
\end{itemize}

\subsection{Optimal Joint Power Allocation (OJPA)}
\label{sec_OJPA}
In this section, we present an optimal approach called the optimal joint power allocation (OJPA) scheme for transmiter and jammer. Since the transitions to state $s^{(k+1)}$ depend solely on the current state $s^{(k)}$ and the current action $a^{(k)}$, our system follows the Markov property. 
Therefore, the proposed problem outlined in (\ref{Problem}) is an online sequential decision-making problem with finite action, state spaces, and a bounded and consistent immediate reward function. We use MDP framework to obtain the solution where the goal is to make optimal decisions at each decision epoch to maximize the expected total reward \cite{puterman2014markov} \cite{bellman1957markovian} \footnote{Our proposed RL approach can be extended to multi-antenna systems, however, the curse of dimensionality, due to increased number of states, is a challenge. Leveraging deep RL techniques offers promising solutions to effectively address these difficulties.}.

In general, a decision rule at the $k$th TS $d^{(k)}$ is expressed as a function of state $s^{(k)}$ such that $a^{(k)}=d^{(k)}(s^{(k)}): \mathcal{S} \xrightarrow{} \mathcal{A}$ denotes the action to be taken at decision epoch $k$ when the system state is $s^{(k)}$. 
Further, a general policy $\pi = \{d^{(0)}(s^{(0)}),d^{(1)}(s^{(1)}),\cdots, d^{(K-1)}(s^{(K-1)})\}$ constitutes a sequence of decision rules in all the decision epochs \cite{puterman2014markov}. The set of all feasible policies is represented by $\Pi$, where $\pi\in\Pi$ should satisfy (\ref{eq_feasible_action}) at all decision epochs. Then, starting with a given state $s^{(0)}$ 
in the first TS and following a policy $\pi$, the expected total reward between the first TS and until the network stops functioning is 
\begin{align}
    V_{\pi}(s^{(0)}) = \mathbb{E}\bigg[\mathbb{E}_{K} \bigg[\sum_{k=0}^{K-1} R^{(k)}(s^{(k)},a^{(k)}) \bigg] \big| s^{(0)},\pi \bigg].
    \label{eq_expected_total_reward}  
\end{align}
Based on the geometric distribution of the lifetime of the network $K$,
(\ref{eq_expected_total_reward}) is equivalent to the expected total discounted reward of an infinite-horizon MDP 
\cite[Proposition~5.3.1]{puterman2014markov} 
\begin{align}
     V_{\pi}(s^{(0)}) = \mathbb{E}\bigg[\sum_{k=0}^{\infty} \Gamma^{k} R^{(k)}(s^{(k)},a^{(k)}) \big| s^{(0)},\pi \bigg],
    \label{eq_objective}    
\end{align} 
where $\Gamma$, the probability that the network remains operational at each TS, can be interpreted as the discount factor of the MDP model \cite{wong2014joint}. Since the network will stop functioning at some time in the future, the reward in the $k$th TS is discounted by a factor $\Gamma^{k}$. The problem in (\ref{eq_objective}) is an infinite-horizon MDP which converges to a finite value\cite[pp.~121]{puterman2014markov}.

We need to find the optimal stationary deterministic policy $\pi^* = \arg \max_{\pi \in \Pi} V_{\pi}(s^{(0)})$ which is the only case of interest in the case of an infinite-horizon MDP by maximizing the expected total discounted reward in (\ref{eq_objective}) \cite{wong2014joint}.  A policy is termed as stationary deterministic when $d^{(k)}(s^{(k)})$ is deterministic (there is a certainty in taking a decision at $s^{(k)}$) Markovian, and $d^{(k)}(s^{(k)}) = d$ for all $k \in K$, resulting in $\pi = (d,d, \cdots)$ \cite[pp.~21]{puterman2014markov}. Hence, the optimal stationary deterministic policy can be denoted as $d^*$. The maximization of the expected total discounted reward in (\ref{eq_objective}) can be implemented by the policy iteration (PI) algorithm \cite[pp.~174]{puterman2014markov}.

The PI algorithm implements \textit{Bellman's equation of optimality} where the optimal expected total discounted reward $V(s)$ for a given current state $s$ is expressed as 
\cite{puterman2014markov}
\begin{align}
V(s)= \max_{a \in U(s)} 
\bigg\{R(s,a) +\Gamma \sum_{s' \in \cal{S}}\mathbb{P}(s'|s,a) V(s')\bigg\}. 
    \label{eq_bellman}
\end{align}
The first term on the right-hand side of (\ref{eq_bellman}) can be interpreted as the immediate reward at the current TS, while the second term signifies the expected total discounted future reward when action $a$ is selected. There exists an optimal stationary deterministic policy $d^*(s)$ which maximizes the right-hand side of (\ref{eq_bellman}) and is given by \cite[Th.~6.2.10]{puterman2014markov} 
\begin{align}
d^*(s)=  \mathop{\mathrm{argmax}}_{a \in U(s)} 
\left\{ R(s,a) + \Gamma \sum_{s' \in \mathcal{S}} \mathbb{P}(s' \mid s,a) V(s') \right\}.
    \label{eq_optimal_policy}    
\end{align}
The pseudo-code for the PI algorithm is given in Algorithm \ref{planning_phase}, which describes steps to finding optimal stationary deterministic policy $d^*(s)$ for each $s \in \mathcal{S}$} and storing these policies in a look-up table. 

We refer to the phase of populating the look-up table in Algorithm \ref{planning_phase} is known as the \textit{planning phase}. The look-up table can then be used at each TS to allocate the power. The planning phase is further divided into two phases, i.e., policy evaluation and policy improvement, as shown in Algorithm \ref{planning_phase}. 
The policy evaluation phase computes $V(s)$ for a given policy $d(s)$ by updating $V(s)$ iteratively from its initial value given in line number one for each $s \in \mathcal{S}$ using the Bellman equation in line number six until it converges in line number nine.
Next, the policy improvement phase finds a better policy $d(s)$ than the given policy $\hat{d}(s)$ for each state  $s \in \mathcal{S}$.
The policy $d(s)$ is obtained by choosing the action $a \in U(s)$ that maximizes $V(s)$ corresponding to $\hat{d}(s)$ in line number thirteen. 
If the policy is unstable ($\hat{d}(s) \neq d(s)$), we repeat the policy evaluation phase again with a better policy obtained in line number thirteen.
The algorithm continues iterating between the policy evaluation and policy improvement phase until the optimal policy is found, when, $\hat{d}(s) = d(s)$.

\begin{algorithm}[t]
\caption{\textbf{Algorithm 1: The Planning Phase}}
\hspace*{\algorithmicindent} \textbf{Input}: Set of states, actions, state transition probability, and reward;\\
\hspace*{\algorithmicindent} \textbf{Output}: Optimal stationary deterministic policy $d^*(s)$ 
\begin{algorithmic}[1]
\State Initialize $V(s)$ and stationary deterministic policy $d(s)$ arbitrarily for all $s\in \mathcal{S}$, set small threshold $\epsilon$.

\begin{flushleft}
\textbf{Policy evaluation:} 
\end{flushleft}
\Repeat
\State $\Delta = 0 $ 
    \ForEach {$s \in \mathcal S $} 
    \State $v = V(s)$
    \State $V(s) =   [R(s,a) + \Gamma \underset{s^{'} \in \mathcal{S}} \sum \mathbb{P}(s^{'} \mid s, a) V(s^{'})]$ 
    \State $\Delta = \max(\Delta, |v- V(s)|)$
    \EndFor
\Until $\Delta < \epsilon $

\begin{flushleft}
 \textbf{Policy improvement:}
 \end{flushleft}
\State policy-stable $=$ true
    \ForEach {$s \in \mathcal S$} 
    \State $\hat{d}(s) = d(s)$
    \State $d(s) = \mathop{\mathrm{argmax}}\limits_{a \in U(s)} 
        \left[ R(s,a) + \Gamma 
        \sum_{s^{'} \in \mathcal{S}} \mathbb{P}(s^{'} \mid s, a) V(s^{'}) \right]$ 
        \If {$\hat{d}(s) \neq d(s)$}
        \State policy-stable $=$ false
        \EndIf
    \EndFor

\begin{flushleft}
\textbf{Check stopping criteria: } 
\end{flushleft}
\If {policy-stable}
\State stop
\Else
\State go-to policy evaluation (line-2) 
\EndIf
\end{algorithmic}
\label{planning_phase}
\end{algorithm}

Next, we refer to the subsequent phase of power allocation as the \textit{transmission phase}. During this phase, the power allocation or action is obtained at each TS by directly fetching the power allocation values corresponding to the states from the look-up table populated by Algorithm \ref{planning_phase}. The pseudo-code of the transmission phase is described in Algorithm \ref{transmission_phase}. \footnote{A CPU in the network which has the look-up table for power allocation from Algorithm \ref{planning_phase} along with global channel states and battery states at the beginning of each TS, can execute Algorithm \ref{transmission_phase} as the decision-maker of the power allocation policies.} In Algorithm \ref{transmission_phase}, we iterate over the decision epochs to obtain the current state $s^{(k)}$ by generating the channel states and battery states first. Then, for each decision epoch, the optimal action $a^{(k)}$ is chosen based on the current state $s^{(k)}$ in line number seven of Algorithm \ref{transmission_phase} from the look-up table created in the planning phase. This provides the joint optimal $P^{(k)}_\textrm{S}$ and $P^{(k)}_\textrm{D}$ for transmission and jamming, respectively, in the $k$th TS. Using these power values, the cumulative reward is obtained and battery states are updated.

\begin{algorithm}[t]
\caption{\textbf{Algorithm 2: Transmission Phase}}
\hspace*{\algorithmicindent} \textbf{Input}: Optimal stationary deterministic policy $d^*(s)$ and initial state $s^{(0)}$\\
\hspace*{\algorithmicindent} \textbf{Output}: Total expected discounted reward defined in (\ref{eqECS})
\begin{algorithmic}[1]
    \State Set $\mu = 0$
    \State Set $k = 0$
    \While{$k \leq K-1$}
    \State Track channel states $G_{\textrm{SD}}^{(k)}, G_{\textrm{SE}}^{(k)}, G_{\textrm{DD}}^{(k)}$ and $G_{\textrm{DE}}^{(k)} $ 
    \State Track available battery $B_\textrm{S}^{(k)}$ and $B_\textrm{D}^{(k)}$ 
    \State Set $s^{(k)} = (G_{\textrm{SD}}^{(k)}, G_{\textrm{SE}}^{(k)}, G_{\textrm{DD}}^{(k)}, G_{\textrm{DE}}^{(k)}, B_\textrm{S}^{(k)}, B_\textrm{D}^{(k)})$
    \State Obtain $a^{(k)} = (P_\textrm{S}^{(k)}, P_\textrm{D}^{(k)})$ from look-up table for state $s^{(k)}$
    \State Consume $P_\textrm{S}^{(k)}$ and $P_\textrm{D}^{(k)}$ for transmission and jamming respectively.
    \State Calculate the total expected discounted reward $\Gamma^{k} C^{(k)}_\textrm{S}$  for state $s^{(k)}$.
    \State Update battery $B_\textrm{S}^{(k)}$ and $B_\textrm{D}^{(k)}$ using (\ref{battery_Bs}) and (\ref{battery_Bd}) respectively
    \State $\mu  = \mu  + \Gamma^{k}C^{(k)}_\textrm{S}  T_s$
    \State Set $k = k + 1$
    \EndWhile
\end{algorithmic}  
\label{transmission_phase}
\end{algorithm}

\subsection{Sub-optimal Joint Power Allocation (SJPA) Algorithms}
Although OJPA algorithm provides optimal performance, it suffers from high computational complexity, which is impractical for sensors nodes having limited computation, storage, and energy resources.
Therefore, in this section, we develop and describe three computationally efficient sub-optimal algorithms.

\subsubsection{ Reduced State Joint Power Allocation (RSJPA)} 
To develop a reduced complexity algorithm, we propose the RSJPA algorithm, which combines the OJPA algorithm outlined in Algorithm \ref{planning_phase} with the GA described in (\ref{GA_algorithm}).
The RSJPA algorithm chooses a smaller subset $\mathcal {S}^\prime \subseteq \mathcal {S}$ by taking states randomly from $\mathcal{S}$ and creates a look-up table for $\mathcal {S}^\prime$ using Algorithm \ref{planning_phase}  in the planning phase \cite{Ajib_VTC}. The reason for randomly selecting a state is that each state in the system has an equal probability. Therefore, choosing any subset of states will not affect the overall performance.
In the transmission phase, the power allocation is carried out from the look-up table for the states in $\mathcal {S}^\prime$ and for the remaining states $\mathcal{S}\setminus\mathcal {S}^\prime$, the GA is applied.
This approach strikes a balance between performance and complexity in comparison to the OJPA algorithm. By increasing the number of states included in the subset $\mathcal{S}^\prime$, we can significantly increase performance; however, this improvement comes with a rise in complexity. This trade-off highlights the importance of carefully considering the state selection to optimize outcomes effectively.
For example, with $N_S/2$ states in $\mathcal {S}^\prime$, we can reduce computation complexity by  $\frac{N_A^{N_S/2}}{2}$ times as compared to that of the OJPA algorithm with $N_S$ states in $\mathcal {S}$.



\subsubsection {Greedy Algorithm (GA)}
The GA algorithm does not require the planning phase.  In the transmission phase, it selects the action $a^{(k)}=\{P^{(k)}_\textrm{S}, P^{(k)}_\textrm{D}\}$ in each TS from the set of feasible actions $U(s^{(k)})$ for the state $s^{(k)}$ that maximizes the immediate reward in (\ref{reward})\cite{Octavia2019}. Accordingly, the power allocation problem is expressed as 
\begin{align}
     a^{(k)} = \mathop{\mathrm{argmax}}\limits_{a^{(k)} \in U(s^{(k)})} R^{(k)}(s^{(k)}, a^{(k)}). 
     \label{GA_algorithm}
\end{align}

\subsubsection {Naive Algorithm (NA)}
The NA algorithm also does not require the planning phase. In the transmission phase, it fully utilizes the energy stored in the battery at node $\textrm{S}$ and node $\textrm{D}$ for transmission and jamming, respectively, in each TS \cite{ahmed2012power} \cite{ahmed2013joint}, i.e., the transmit and jamming power in the $k$th TS are $P^{(k)}_\textrm{S} = \frac{B^{(k)}_\textrm{S}}{T_s}$ and $P^{(k)}_\textrm{D} = \frac{B^{(k)}_\textrm{D}}{ T_s}$, respectively.

Implementing the algorithms proposed in this paper requires global channel state information, which includes the channel state information related to the eavesdropper. In certain scenarios, it may be possible to acquire the eavesdropper's channel state information when the eavesdropper is an active node in the network and its transmissions can be monitored \cite{poor2009secrecy, poor2008secure}.
For instance, in networks where nodes serve dual roles, acting as legitimate receivers for some transmissions while functioning as eavesdroppers for others, the channel state information of the eavesdroppers may be obtained. Another example is found in networks where confidential information is intended solely for a specific user, treating all other nodes as potential eavesdroppers, as seen in military communications. In such cases, any data transmission from the eavesdropper to the source and destination could enable the estimation of the eavesdropper’s channel state information by leveraging the reciprocal characteristics of the wireless channel.
Furthermore, in the secure communication literature, it is a common assumption that channel state information related to eavesdroppers is available \cite{insoo2019FD, hoseini2023GC, yang2020_PLS_drl_TWC, saleem2022_PLS_drl, liu2024secrecy_DRL, li2022DRL, yang2024EH_DRL, qian2022secrecy, poor2009secrecy, poor2008secure, poor2014security, kundu2021ergodic, shashi2024TVT}.

For the practical implementation of the joint power allocation algorithms OJPA and RSJPA, a CPU with access to all the system information, fed back from both the source and destination, can generate a look-up table for power allocation. The CPU can then instruct the source and destination to configure their respective power levels at each TS based on the look-up table corresponding to the system states.

\section{Individual Power Allocation (IPA)}

In this section, we now consider systems with a single EH node where either node $\textrm{S}$ or node $\textrm{D}$ is EH.
The transmit power of the single EH node system is optimized while the other node relies on a fixed power supply. When the transmit power of node $\textrm{S}$ is optimized with a fixed power supply at node $\textrm{D}$, we refer to this case as individual transmit power allocation. When the jamming power of node $\textrm{D}$ is optimized while the fixed power supply is at node $\textrm{S}$, we refer to this case as individual jamming power allocation. We apply the optimal and sub-optimal solution strategies (modified accordingly) described in Section \ref{section_prposed_solution} for computing the power allocation in these cases.

\subsection {Individual Transmit Power Allocation (ITPA)} 
\label{sec_transmit_power_optimization}
 In this case, we only optimize $P_\textrm{S}^{(k)}$ when
 $P_\textrm{D}^{(k)} = P_\textrm{D}$ for each TS. 
To this end, we modify the MDP accordingly. The state and action set of the system in the $k$th TS can now be represented as $s^{(k)} = (G_{\textrm{SD}}^{(k)}, G_{\textrm{SE}}^{(k)}, G_{\textrm{DD}}^{(k)}, G_{\textrm{DE}}^{(k)}, B_\textrm{S}^{(k)})$ and $a^{(k)} \in U(s^{(k)}) = \big\{P_\textrm{S}^{(k)} \mid 0 \leq P_\textrm{S}^{(k)} \leq \frac{B_{\textrm{S}}^{(k)}}{T_s}\big\}$,
respectively. The transition probabilities change based on $s^{(k)}$ and $a^{(k)}$ by following (\ref{Transition_Prob}). With these changes, to obtain optimal solution, we apply Algorithm \ref{planning_phase} and   Algorithm \ref{transmission_phase} and to obtain sub-optimal solution, we apply NA, GA and RSJPA algorithms.
    
\subsection {Individual Jamming Power Allocation (IJPA)} 
\label{sec_jamming_power_optimization}
 In this case, we only optimize $P_\textrm{D}^{(k)}$ when
 $P_\textrm{S}^{(k)} = P_\textrm{S}$ for each TS.  
Following changes in section \ref{sec_transmit_power_optimization}, the state and action set of the system in the $k$th TS can be represented as  $s^{(k)} = (G_{\textrm{SD}}^{(k)}, G_{\textrm{SE}}^{(k)}, G_{\textrm{DD}}^{(k)}, G_{\textrm{DE}}^{(k)},  B_\textrm{D}^{(k)})$ and $a^{(k)} \in U(s^{(k)}) = \big\{P_\textrm{D}^{(k)} \mid 0 \leq P_\textrm{D}^{(k)} \leq \frac{B_{\textrm{D}}^{(k)}}{T_s}\big\}$, respectively. The transition probabilities will change based on $s^{(k)}$ and $a^{(k)}$ by following (\ref{Transition_Prob}). With these changes, optimal and sub-optimal algorithms are applied as in section \ref{section_prposed_solution}.





\section{Computational Complexity Analysis}
\label{sec_complexity}

In this section, we analyze the computational complexity of aforementioned algorithms, i.e., OJPA, RSJPA, GA and NA.


In the worst case, Algorithm \ref{planning_phase} of the OJPA algorithm may need to consider all possible policies. 
For each state, there are $N_A$ actions to choose from. As there are $N_S$ states, the total number of possible policies is $N_A^{N_S}$. As $N_A$ and $N_S$ grow, the number of possible policies grows exponentially.
During each iteration of policy improvement, the algorithm typically eliminates several sub-optimal policies. 
As each iteration of policy improvement reduces the remaining policy space, on average, the worst-case computational complexity of the planning phase of the OJPA algorithm is $\mathcal{O}\big(\frac{N_A^{N_S}}{N_S}\big)$ \cite{PIcomplexity}. 
In the transmission phase, power allocation at each TS is implemented by directly fetching power allocation values from the look-up table for that TS. As there are $K$ TSs, the transmission phase complexity is  $\mathcal{O}(K)$.

The complexity of the RSJPA algorithm in the planning phase with $\rho \%$ of $N_S$ states in $\mathcal{S}^\prime$,  is given by $\mathcal{O}\big(\frac{N_A^{{\rho N_S }/{100}}}{{\rho N_S}/{100}}\big)$. This is because the planning phase of the RSJPA is implemented by executing the OJPA algorithm with ${\rho N_S}/{100}$ states \cite{Ajib_VTC}.
When it comes to transmission phase complexity, the worst case possibility is that none of the states in the transmission phase belong to the look-up table due to the arbitrary selection of states for the preparation of the look-up table in the planning phase. That is why the worst case complexity in the transmission phase is $\mathcal{O}(KN_A)$ as for each TS, GA algorithm is implemented. On the contrary, the best case possibility is that all the states in the transmission phase belong to the look-up table. In this case, the complexity would be $\mathcal{O}(K)$ as in the OJPA algorithm. 
In the average case, states for the $\rho \%$ of the TSs might belong to the look-up table but the states for the remaining $(100-\rho) \%$ TSs might not. In this case, the average complexity would be $\mathcal{O}(\frac{K}{100}(\rho + (100-\rho)N_A))$.

The GA does not require a planning phase. Its transmission phase complexity is $\mathcal{O}(K N_A)$ as $N_A$ computations are required to identify the best action among $N_A$ actions that maximize the current reward at each TS \cite{Ajib_VTC}.
As in the GA, the NA also does not require a planning phase. As we just use the maximum stored energy in the batteries for transmission in each TS, the transmission phase complexity of the NA is $\mathcal{O}(K)$ \cite{ahmed2013joint}.

\begin{table}
\centering
\begin{tabular}{|p{1.8cm}|p{2cm}|p{3.8cm}| }
 \hline
 \textbf{Algorithms}  & \textbf{Planning phase} & \textbf{Transmission phase}\\ 
 \hline
  OJPA &  $\mathcal{O}\big(\frac{N_A^{N_S}}{N_S}\big)$ & $\mathcal{O}(K)$\\ 
  \hline
  & & \\
  RSJPA & $ \mathcal{O}\big(\frac{N_A^{{\rho N_S }/{100}}}{{\rho N_S}/{100}}\big)$ 
  & Best case: $\mathcal{O}(K)$\\
   & & Average case: $\mathcal{O}(\frac{K}{100}(\rho + (100-\rho)N_A))$ \\
    & & Worst case: $\mathcal{O}(KN_A)$\\
  \hline
  GA & $-$ & $\mathcal{O}(KN_A)$ \\
  \hline
  NA & $-$ & $\mathcal{O}(K)$ \\
  \hline

 \end{tabular}
\caption{Complexities of different algorithms.}
 \label{complexity}
\end{table}

\section{Results and Discussions}

\begin{table*}
    \centering
    \begin{tabular}{  p{7.8cm}|c| c p{6.3cm} |  }
         \hline
         \textbf{Description} &  \textbf{Notation} &  \textbf{Value}\\
         \hline
         Channel bandwidth   & $W$   & $2$ MHz \\
         \hline
        Noise power spectral density  & $N_0$  & $10^{-20.4}$ W/Hz   \\
        \hline
        Channel power gain set & $\mathcal{G}$   & $\{G_1, G_2\} = \{1.655 \times 10^{-13}, 3.311 \times 10^{-13}\}$\\
        \hline
        Channel state transition probability matrix    & $\mathbb{P}(G_{\textrm{XY}}^{(k+1)}\mid G_{\textrm{XY}}^{(k)})$
         &  
        $\begin{bmatrix}
        0.9 & 0.1\\
        0.1 & 0.9 
        \end{bmatrix}$\\
        \hline
        Self-interference coefficient  & $\alpha$  & $10^{-5}$ \\
        \hline
        Stopping criteria for policy evaluation loop  & $\epsilon$   & $0.07$ \\
        \hline
        Duration of TS & $T_s$ & 5ms\\
        \hline
        One energy unit & & $2.5$ $\mu$J \\
        \hline
        Harvested energy   & \{$E_\textrm{S}$, $E_\textrm{D}$\}  & \{$1$, $2$\}, \{$2$, $1$\} energy units\\
        \hline
        Battery capacity of node $\textrm{S}$ & $B_{\textrm{S}}^{\textrm{max}}$ & $5$ energy units\\
        \hline
        Battery capacity of node $\textrm{D}$ & $B_{\textrm{D}}^{\textrm{max}}$ & $5$ energy units\\
        \hline
        Probability of harvesting $E_\textrm{S}$ units of energy at $\textrm{S}$ & $p$ &  $0.5, 0.8$\\
        \hline
        Probability of harvesting $E_\textrm{D}$ units of energy at $\textrm{D}$ & $q$ & $0.5, 0.8$\\
        \hline
        Set of transmit and jamming power  & $\mathcal{P}$ & $\{0,0.5,1,2\}$ mW  \\
        \hline
        Set of transmit and jamming energy & $\mathcal{E_U}$ &  $\{0,1,2,4\}$ energy unit\\
        \hline

        Initial state & $s^{(1)}$ & $(G_2,G_2,G_2,G_2, B_{\textrm{S}}^{\textrm{max}},B_{\textrm{D}}^{\textrm{max}})$ \\
         \hline 
    \end{tabular}
    \caption{Simulation parameters}
    \label{simu_para}
\end{table*}

In this section, we compare the performance of OJPA and SJPA (RSJPA, GA, and NA) algorithms in terms of expected total discounted reward (expected total transmitted secure bits) and energy efficiency. We also compare the energy efficiency of the OJPA algorithm with that of the two IPA algorithms ITPA and IJPA.  
The energy efficiency $\eta_{E}$ of a network in bits per energy unit is defined as the expected ratio of the total transmitted secure bits and the total transmitted energy until the network stops functioning 

\begin{align}
        \eta_{E}
    =  \mathbb{E}  \left[\mathbb{E}_K  \left[ 
        \frac{\sum_{k=0}^{K-1}   C_\textrm{S}^{(k)} T_s}
    {\sum_{k=0}^{K-1}(P_\textrm{S}^{(k)}+P_\textrm{D}^{(k)})T_s}\right]\right].
        \label{EE}
    \end{align}

The energy efficiency is evaluated by using the $P_\textrm{S}$ and $P_\textrm{D}$ obtained from the proposed OJPA and SJPA (RSJPA, GA, and NA) algorithms.
We use a typical personal computer with an Intel\textsuperscript{\textregistered} Core\textsuperscript{TM} i7-8700 CPU and
16 GB RAM to implement the algorithms.
The list of simulation parameters with their values is given in Table \ref{simu_para}. The values of the parameters are mostly taken from \cite{dohler2013learning, ieee1997wireless}. 


\begin{figure}\centering

     \begin{subfigure}[t]{0.24\textwidth}
     \fontsize{15}{12}\selectfont 
         \centering
         \includegraphics[width=\textwidth]{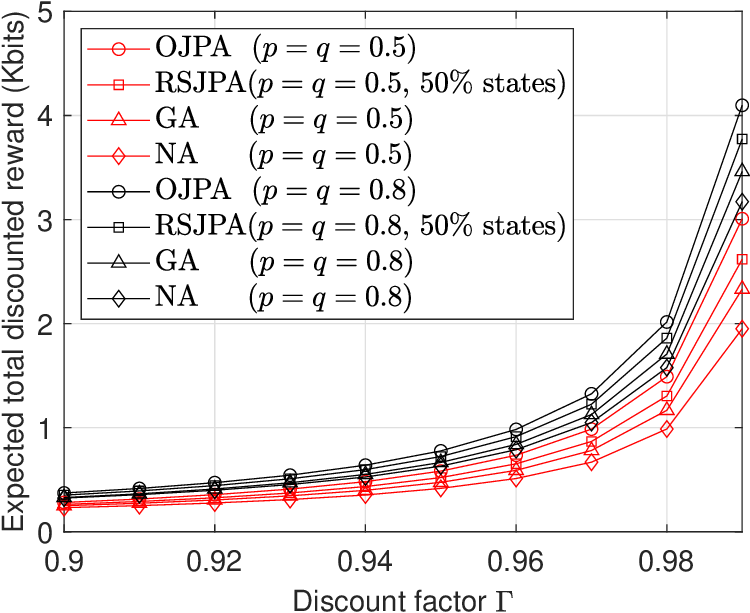}
         \caption{Expected total discounted reward versus $\Gamma$ for joint power allocation algorithms for different $p$ and $q$ with $B_{\textrm{S}}^{\textrm{max}} = B_{\textrm{D}}^{\textrm{max}} = 5$ energy units. }
         \label{fig_ETD_vs_Gamma}
         \vspace{0.3cm}
     \end{subfigure}
      \hfill
    \begin{subfigure}[t]{0.24\textwidth}
         \centering
         \includegraphics[width=\textwidth]{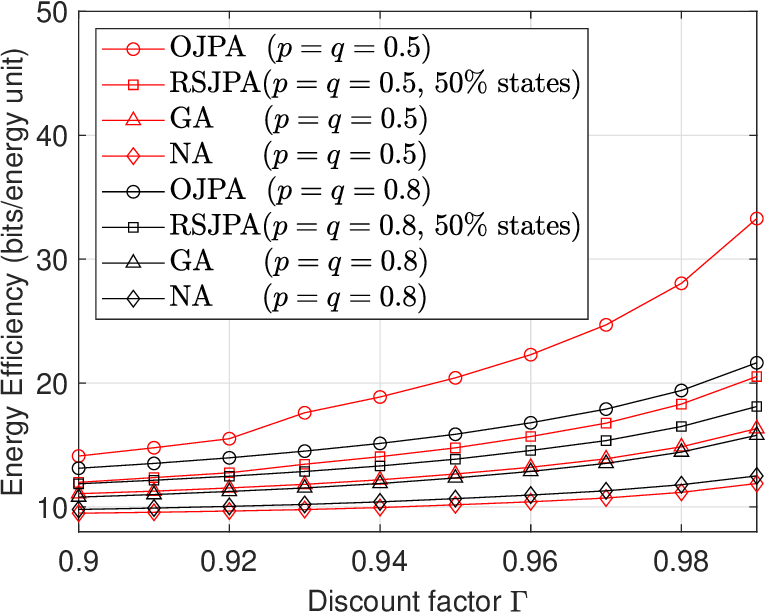}
         \caption{Energy efficiency versus $\Gamma$ for joint power allocation algorithms for different $p$ and $q$ with $B_{\textrm{S}}^{\textrm{max}} = B_{\textrm{D}}^{\textrm{max}} = 5$ energy units. }
         \label{fig_EE_vs_Gamma}
     \end{subfigure}
 
     \begin{subfigure}[t]{0.24\textwidth}
     \fontsize{15}{12}\selectfont 
         \centering
         \includegraphics[width=\textwidth]{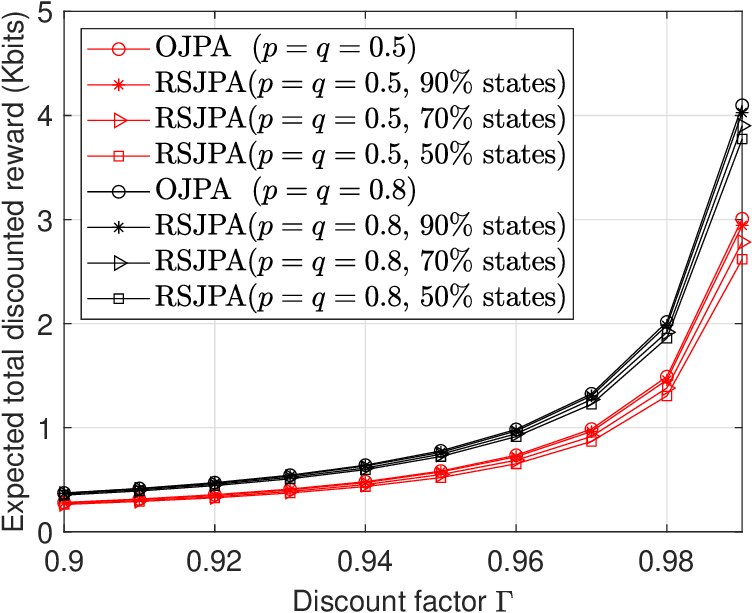}
         \caption{Expected total discounted reward versus $\Gamma$ for joint power allocation algorithms for different $p$ and $q$ with $B_{\textrm{S}}^{\textrm{max}} = B_{\textrm{D}}^{\textrm{max}} = 5$ energy units.}
         \label{fig_ETD_vs_Gamma_varyRSJPA}
         \vspace{0.3cm}
     \end{subfigure}
         \hfill
    \begin{subfigure}[t]{0.24\textwidth}
         \centering
         \includegraphics[width=\textwidth]{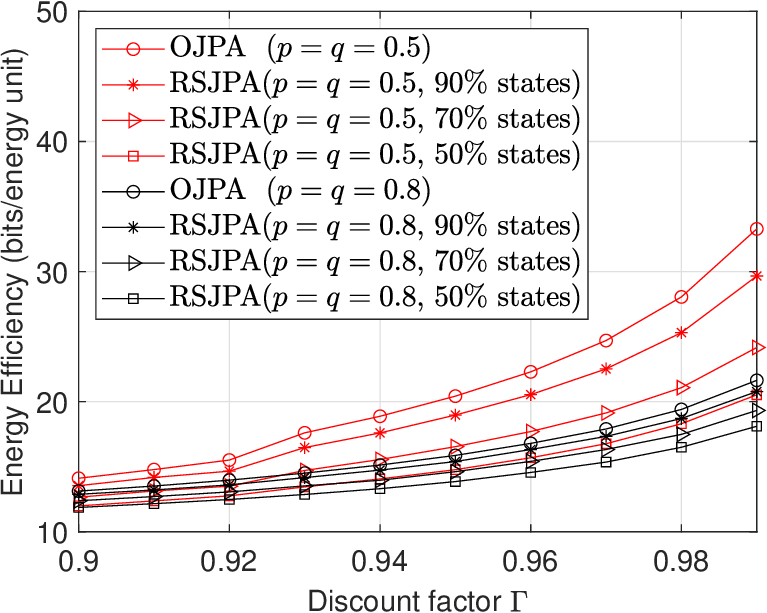}
         \caption{Energy efficiency versus $\Gamma$ for joint power allocation algorithms for different $p$ and $q$ with $B_{\textrm{S}}^{\textrm{max}} = B_{\textrm{D}}^{\textrm{max}} = 5$ energy units.}
         \label{fig_EE_vs_Gamma_varyRSJPA}
     \end{subfigure}

     \caption{Expected total discounted reward and energy efficiency versus discount factor $\Gamma$.}
     \label{fig_ETD_EE_vs_Gamma_Joint}
     \vspace{-0.4cm}
     \normalsize 
\end{figure}

In Fig. \ref{fig_ETD_vs_Gamma} we compare the expected total discounted reward versus $\Gamma$ for the algorithms OJPA, RSJPA, GA, and NA when the probability of EH improves from $p=q=0.5$ to $p=q=0.8$ while harvested energy $E_\textrm{S}=E_\textrm{D}=2$. 
The corresponding energy efficiency  plot is shown in Fig. \ref{fig_EE_vs_Gamma}. 
The motivations behind plotting Fig. \ref{fig_ETD_vs_Gamma} and  Fig. \ref{fig_EE_vs_Gamma} are four-fold. First is to check whether a higher value of $\Gamma$ leads to a larger expected total discounted reward and energy efficiency or not.  As $1/(1-\Gamma)$ is the average lifetime of the network, a higher value of $\Gamma$ indicates a longer lifetime. Thus, a higher value of $\Gamma$ should lead to a better performance. The second is to assess the relative performance of algorithms, OJPA, RSJPA, GA, and NA. The third is to find how EH probability $p$ and $q$ affect performance when they are equal. Lastly, is to study the impact of varying the number of states in $\mathcal{S}^{\prime}$ on the performance of RSJPA.

It can be observed from Figs. \ref{fig_ETD_vs_Gamma} - \ref{fig_EE_vs_Gamma_varyRSJPA}  that the higher value of $\Gamma$ leads to better expected total discounted reward and energy efficiency.
From Fig. \ref{fig_ETD_vs_Gamma}, it can be observed that the OJPA algorithm performs the best and the NA performs the worst as expected. The OJPA algorithm takes into account the long-term performance of the system and that is why its performance is the best. The NA performs the worst because it does not consider both immediate and future rewards. Instead, it simply utilizes all the stored energy in the battery at every TS. The GA outperforms the NA as it prioritizes maximizing immediate reward. 
The performance of the RSJPA algorithm is close to the OJPA algorithm being better than the GA which is near the OJPA algorithm as it adopts a hybrid approach between the OJPA algorithm and the GA.

\begin{figure}\centering

     \begin{subfigure}[t]{0.24\textwidth}
     \fontsize{15}{12}\selectfont 
         \centering
         \includegraphics[width=\textwidth]{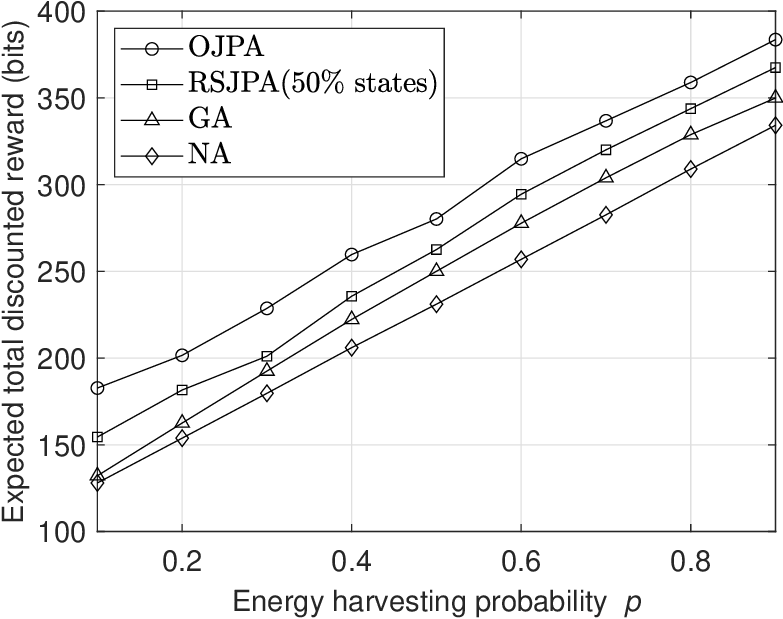}
         \caption{When $q=0.5$, $B_{\textrm{S}}^{\textrm{max}}= B_{\textrm{D}}^{\textrm{max}} = 5$ energy units and $\Gamma = 0.9$. }
         \label{fig_ETD_vs_EH_varyp}
         \vspace{0.3cm}
     \end{subfigure}
      \hfill
     \begin{subfigure}[t]{0.24\textwidth}
         \centering
         \includegraphics[width=\textwidth]{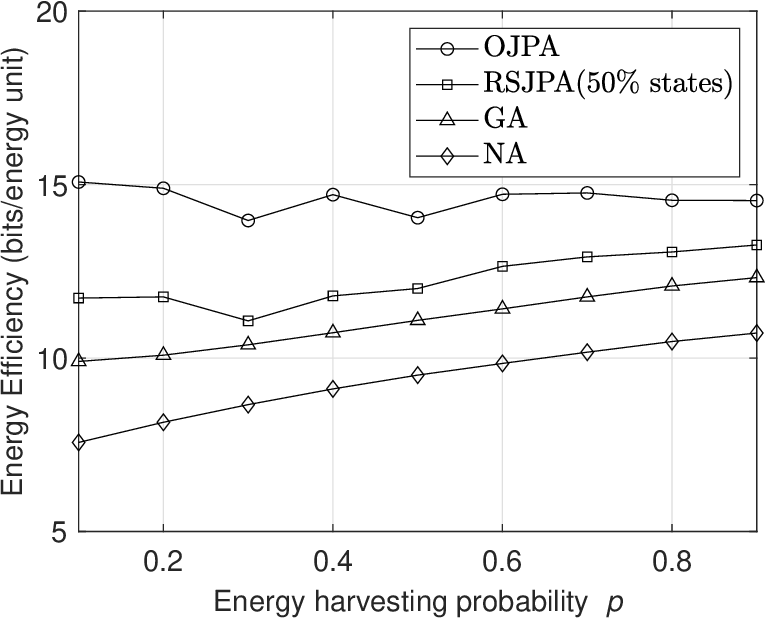}
         \caption{When $q=0.5$, $B_{\textrm{S}}^{\textrm{max}}= B_{\textrm{D}}^{\textrm{max}} = 5$ energy units and $\Gamma = 0.9$.}
         \label{fig_EE_vs_EH_varyp}
         \vspace{0.3cm}
     \end{subfigure}    
     
    \begin{subfigure}[t]{0.24\textwidth}
         \centering
         \includegraphics[width=\textwidth]{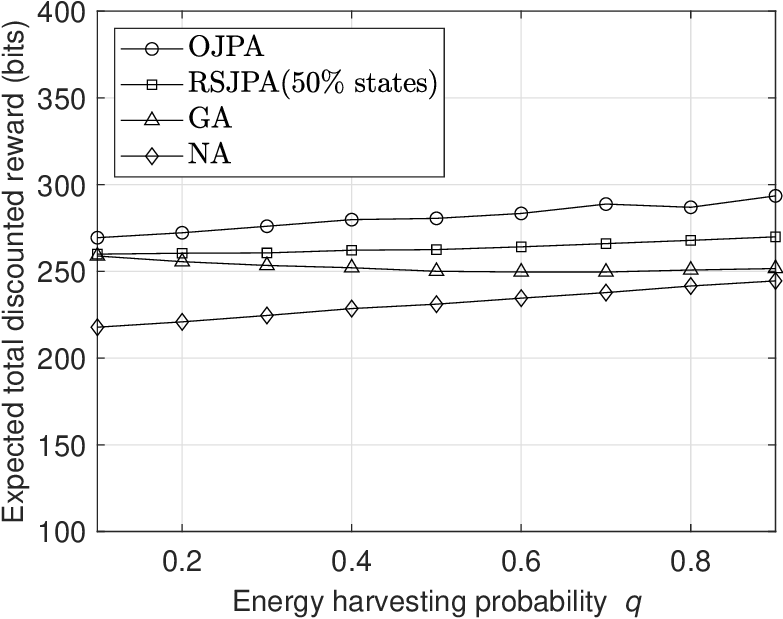}
         \caption{When $p=0.5$, $B_{\textrm{S}}^{\textrm{max}}= B_{\textrm{D}}^{\textrm{max}} = 5$ energy units and $\Gamma = 0.9$. }
         \label{fig_ETD_vs_EH_varyq}
     \end{subfigure}
        \hfill
    \begin{subfigure}[t]{0.24\textwidth}
         \centering
         \includegraphics[width=\textwidth]{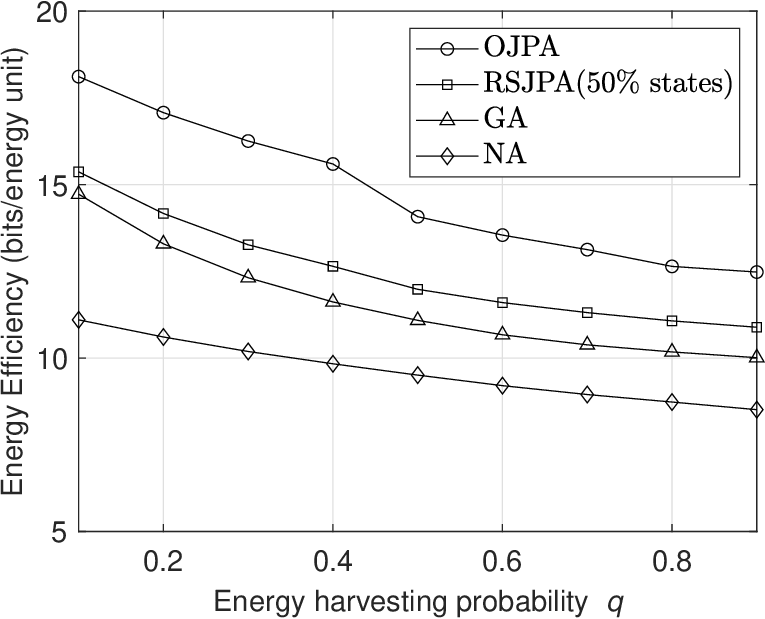}
         \caption{When $p=0.5$, $B_{\textrm{S}}^{\textrm{max}}= B_{\textrm{D}}^{\textrm{max}} = 5$ energy units and $\Gamma = 0.9$.}
         \label{fig_EE_vs_EH_varyq}
     \end{subfigure}
 
     \caption{Expected total discounted reward and energy efficiency versus EH probability $p$ and $q$.}
     \label{fig_ETD_EE_vs_EH_Joint_varypq}
     \vspace{-0.4cm}
     \normalsize 
\end{figure}

We observe from Fig. \ref{fig_ETD_vs_Gamma} that as $p$ and $q$ improve, all the algorithms tend to perform better in terms of the expected total discounted reward.  
In contrast, we notice from Fig. \ref{fig_EE_vs_Gamma} that the energy efficiency for all the algorithms is better when the probability of EH decreases except for the NA. 
This observation prompts us to plot Fig. \ref{fig_ETD_EE_vs_EH_Joint_varypq} to closely study how the expected total discounted reward and energy efficiency vary with the probability of EH and the reasoning for the same. 
A common observation from both Figs. \ref{fig_ETD_vs_Gamma} and \ref{fig_EE_vs_Gamma} is that the performance gap between the OJPA algorithm and other algorithms is greater when the probability of EH decreases. This suggests that the OJPA algorithm is more beneficial at a lower probability of EH.

Also, Fig. \ref{fig_ETD_vs_Gamma} and Fig. \ref{fig_EE_vs_Gamma}  show the performance of the RSJPA algorithm where the algorithm is executed with only $50\%$ of the total number of system states. 
In Fig. \ref{fig_ETD_vs_Gamma_varyRSJPA} and Fig. \ref{fig_EE_vs_Gamma_varyRSJPA}, we consider the same metrics as it is in for Fig. \ref{fig_ETD_vs_Gamma} and Fig. \ref{fig_EE_vs_Gamma}, respectively, with varying number of state  (i.e., $50\%$ to $90\%$) for RSJPA algorithm. Only the comparison between the OJPA and RSJPA is shown. 
The performance gap between the OJPA and RSJPA algorithms in both Fig. \ref{fig_ETD_vs_Gamma_varyRSJPA} and Fig. \ref{fig_EE_vs_Gamma_varyRSJPA} diminishes as the number of states selected for the RSJPA algorithm execution increases.
Thus, the performance of the RSJPA algorithm gradually converges to that of the OJPA algorithm as the number of states in the RSJPA increases. 
However, it should also be noted that the complexity of the RSJPA algorithm also gradually tends towards that of the OJPA algorithm.

\begin{figure}\centering

     \begin{subfigure}[t]{0.24\textwidth}
     \fontsize{15}{12}\selectfont 
         \centering
         \includegraphics[width=\textwidth]{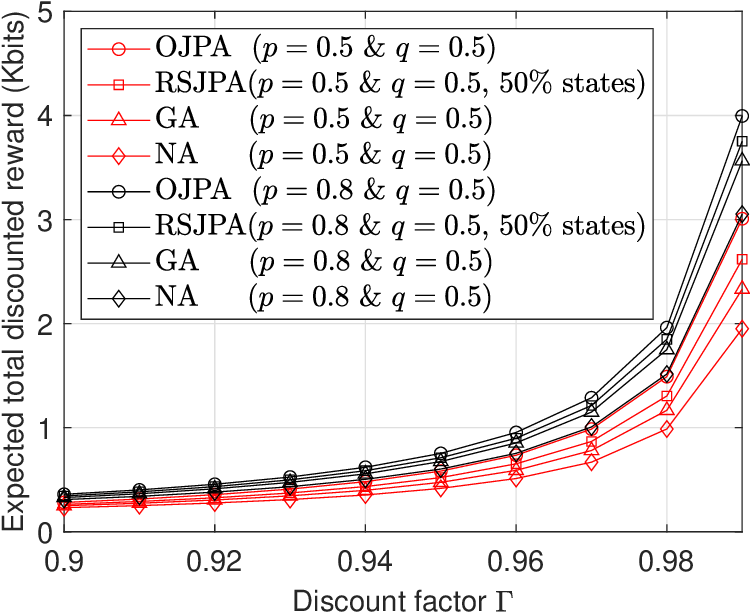}
         \caption{When $p$ improves from $0.5$ to $0.8$ with $q=0.5$ and $B_{\textrm{S}}^{\textrm{max}} = B_{\textrm{D}}^{\textrm{max}} = 5$ energy units. }
         \label{fig_ETD_varyP}
         \vspace{0.3cm}
     \end{subfigure}
      \hfill
    \begin{subfigure}[t]{0.24\textwidth}
         \centering
         \includegraphics[width=\textwidth]{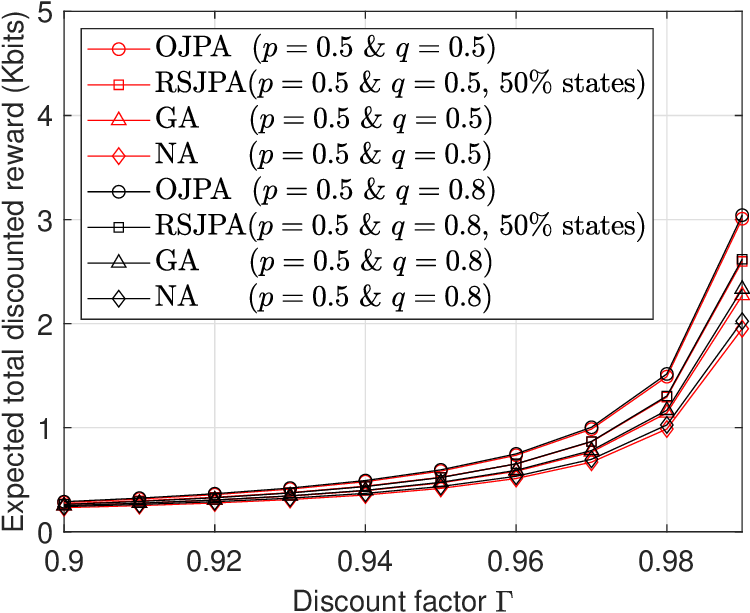}
         \caption{When $q$ improves from $0.5$ to $0.8$ with $p=0.5$ and $B_{\textrm{S}}^{\textrm{max}} = B_{\textrm{D}}^{\textrm{max}} = 5$ energy units.}
         \label{fig_ETD_varyQ}
     \end{subfigure}
 
     \begin{subfigure}[t]{0.24\textwidth}
     \fontsize{15}{12}\selectfont 
         \centering
         \includegraphics[width=\textwidth]{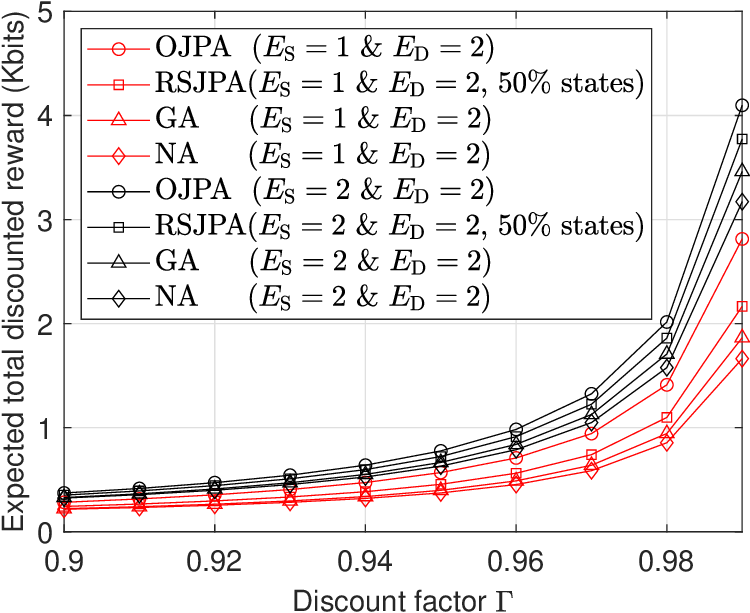}
         \caption{When $E_\textrm{S}$ improves from $1$ to $2$ with $E_\textrm{D}=2$ and $B_{\textrm{S}}^{\textrm{max}} = B_{\textrm{D}}^{\textrm{max}} = 5$ energy units.}
         \label{fig_ETD_varyS}
         \vspace{0.3cm}
     \end{subfigure}
         \hfill
    \begin{subfigure}[t]{0.24\textwidth}
         \centering
         \includegraphics[width=\textwidth]{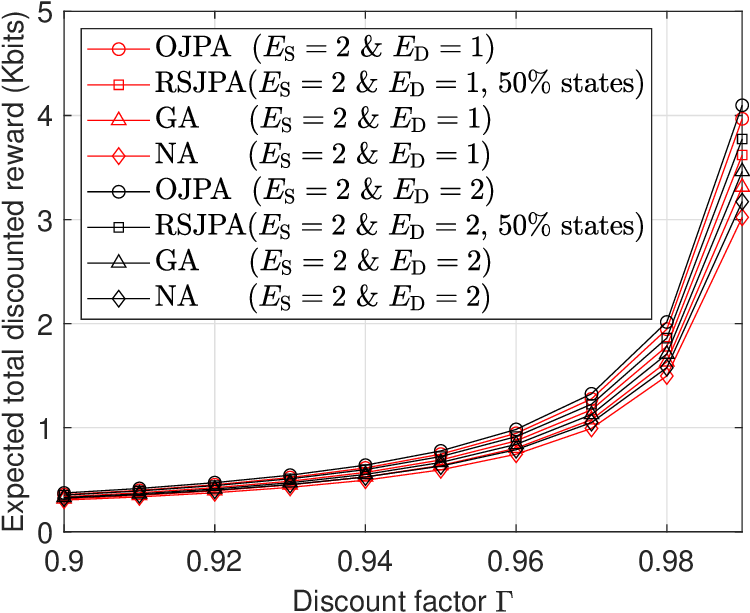}
         \caption{When $E_\textrm{D}$ improves from $1$ to $2$ with $E_\textrm{S}=2$ and $B_{\textrm{S}}^{\textrm{max}} = B_{\textrm{D}}^{\textrm{max}} = 5$ energy units.}
         \label{fig_ETD_varyD}
     \end{subfigure}

     \caption{Expected total discounted reward versus discount factor $\Gamma$ with unequal probability of EH and harvested energy units at $\textrm{S}$ and $\textrm{D}$.}
     \label{fig_ETD_vs_Gamma_P&Q}
     \vspace{-0.4cm}
     \normalsize 
\end{figure}

Figs. \ref{fig_ETD_varyP} and \ref{fig_ETD_varyQ} depict the same performance metric as of Fig. \ref{fig_ETD_vs_Gamma} when $p$ and $q$ are unequal keeping $E_\textrm{S}=E_\textrm{D}=2$, and Figs. \ref{fig_ETD_varyS} and \ref{fig_ETD_varyD} depict the same considering $E_\textrm{S}$ and $E_\textrm{D}$ to be unequal, when $p=q=0.8$.
In Fig. \ref{fig_ETD_varyP}, $p$ increases from $0.5$ to $0.8$ when $q=0.5$, whereas in Fig. \ref{fig_ETD_varyQ}, $q$ increases from $0.5$ to $0.8$ when $p=0.5$. In both figures, increasing EH probability either at $\textrm{S}$ or $\textrm{D}$ improves performance. However, the performance improvement is greater when EH probability improves at $\textrm{S}$. 
The observation in Figs. \ref{fig_ETD_varyS} and \ref{fig_ETD_varyD} with the change in harvested energy is similar, that is, when harvested energy increases from 1 to 2 energy units at $\textrm{S}$ rather than at $\textrm{D}$, the performance improvement is more. We can conclude from Fig. \ref{fig_ETD_vs_Gamma_P&Q} that the improvement of the probability of EH and the harvested energy at both $\textrm{S}$ and $\textrm{D}$ is beneficial for the system; however, the improvement of these metrics at $\textrm{S}$ has a greater impact on the improvement of the performance of the system.

Fig. \ref{fig_ETD_vs_EH_varyp} plots the expected total discounted reward versus the EH probability $p$ while $q=0.5$ and compares the performance of algorithms OJPA, RSJPA, GA, and NA. 
The corresponding energy efficiency plot can be found in Fig. \ref{fig_EE_vs_EH_varyp}. 
The same plots of Fig. \ref{fig_ETD_vs_EH_varyp} and Fig. \ref{fig_EE_vs_EH_varyp} are replicated in Fig. \ref{fig_ETD_vs_EH_varyq} and Fig. \ref{fig_EE_vs_EH_varyq}, respectively, when the EH probability $q$ is varied while $p=0.5$.
From  Fig. \ref{fig_ETD_vs_EH_varyp} we find that the expected total discounted reward increases for all of the algorithms as EH probability $p$ increases. However, in Fig. \ref{fig_ETD_vs_EH_varyq}, the expected total discounted reward increases for all of the algorithms except for the GA algorithm as EH probability $q$ increases. A higher probability of EH leads to more energy available at $\textrm{S}$ or $\textrm{D}$ which leads to more expected total transmitted secure bits until the network stops functioning, hence, the observation except for the GA in Fig. \ref{fig_ETD_vs_EH_varyq}.  We also observe that the rate of performance improvement with the EH probability is higher in Fig. \ref{fig_ETD_vs_EH_varyp} when $p$ increases at $\textrm{S}$ as compared to Fig. \ref{fig_ETD_vs_EH_varyq} when $q$ increases at $\textrm{D}$.

More expected total transmitted secure bits do not mean more energy efficiency. Energy efficiency decreases with the increasing EH probability $p$ in Fig. \ref{fig_EE_vs_EH_varyp} for the OJPA algorithm and all the algorithms in  Fig. \ref{fig_EE_vs_EH_varyq} with the increasing EH probability $q$. However, for the RSJPA, GA, and NA algorithms, the energy efficiency increases in Fig. \ref{fig_EE_vs_EH_varyp}. 
As energy efficiency is the ratio of total transmitted secure bits and total energy expenditure in the system, it seems for the algorithms the rate of increase in the total transmitted secure bits and the rate of increase in the energy expenditure are not the same in various parameter combinations.
This suggests that the optimization for total transmitted secure bits alone without taking energy efficiency into account is not beneficial. Rather one needs to optimize the system taking both into account. 

\begin{figure}\centering

     \begin{subfigure}[t]{0.24\textwidth}
     \fontsize{15}{12}\selectfont 
         \centering
         \includegraphics[width=\textwidth]{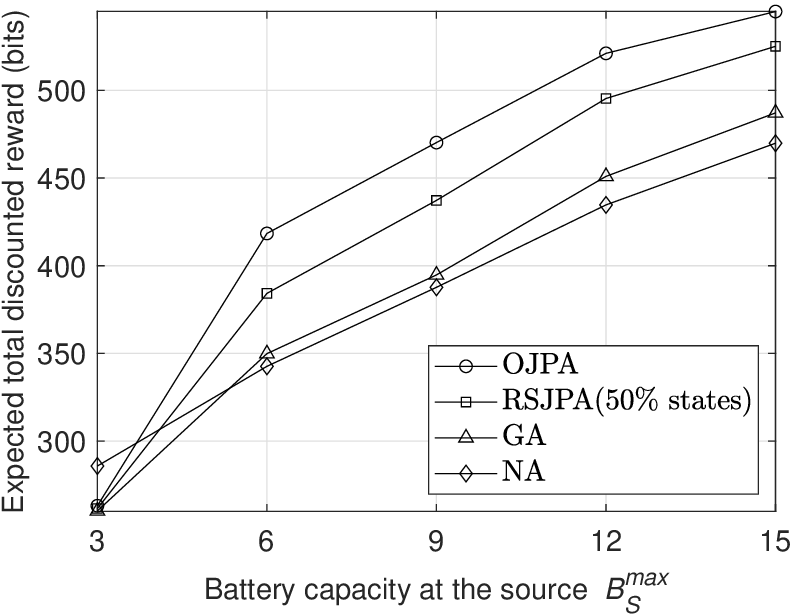}
         \caption{Expected total discounted reward versus $B_{\textrm{S}}^{\textrm{max}}$ for joint power allocation algorithms where $B_{\textrm{D}}^{\textrm{max}} = 5$ energy units. }
         \label{fig_ETD_vs_Bsmax}
         \vspace{0.3cm}
     \end{subfigure}
      \hfill
     \begin{subfigure}[t]{0.24\textwidth}
         \centering
         \includegraphics[width=\textwidth]{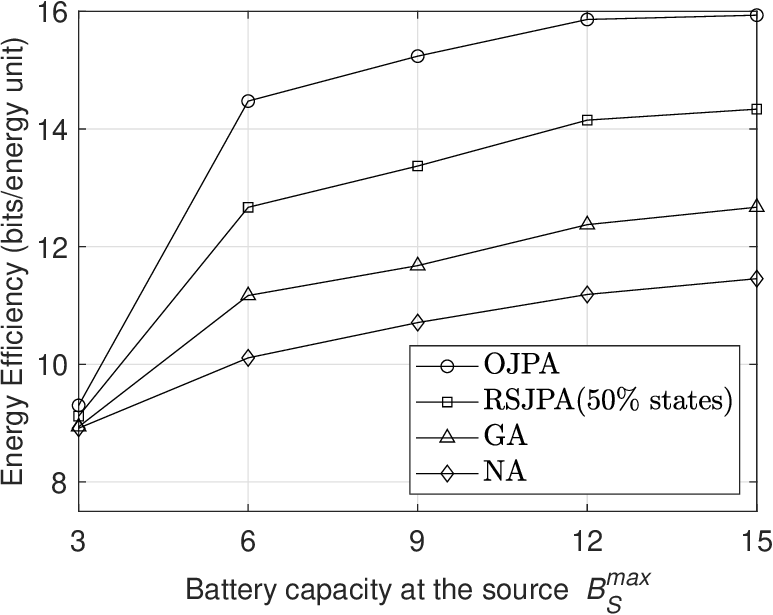}
         \caption{Energy efficiency versus $B_{\textrm{S}}^{\textrm{max}}$ for joint power allocation algorithms where $B_{\textrm{D}}^{\textrm{max}} = 5$ energy units.}
         \label{fig_EE_vs_Bsmax}
         \vspace{0.3cm}
     \end{subfigure}    
     
    \begin{subfigure}[t]{0.24\textwidth}
         \centering
         \includegraphics[width=\textwidth]{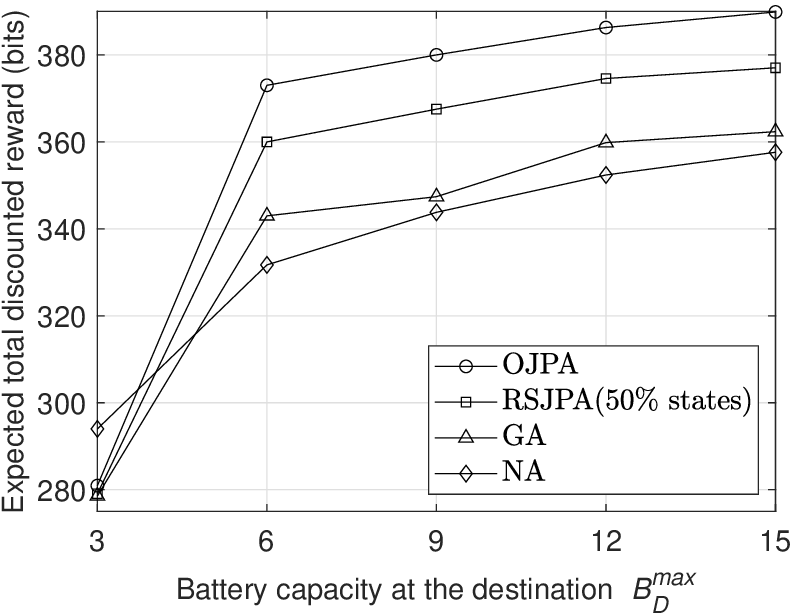}
         \caption{Expected total discounted reward versus $B_{\textrm{D}}^{\textrm{max}}$ for joint power allocation algorithms where $B_{\textrm{S}}^{\textrm{max}} = 5$ energy units.}
         \label{fig_ETD_vs_Bdmax}
     \end{subfigure}
        \hfill
    \begin{subfigure}[t]{0.24\textwidth}
         \centering
         \includegraphics[width=\textwidth]{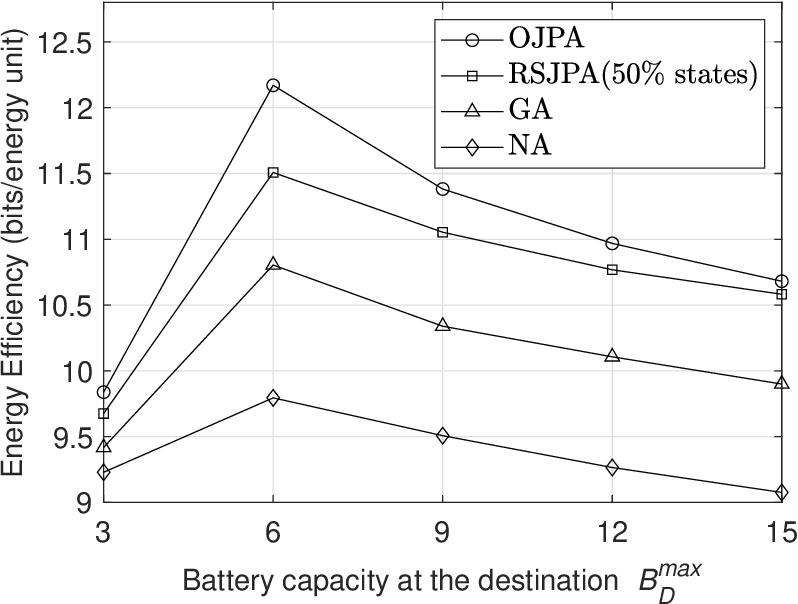}
         \caption{Energy efficiency versus $B_{\textrm{D}}^{\textrm{max}}$ for joint power allocation algorithms where $B_{\textrm{S}}^{\textrm{max}} = 5$ energy units.}
         \label{fig_EE_vs_Bdmax}
     \end{subfigure}
 
     \caption{Expected total discounted reward and energy efficiency versus battery capacity $B_{\textrm{S}}^{\textrm{max}}$ and $B_{\textrm{D}}^{\textrm{max}}$ for joint power allocation algorithms where, $p=q=0.8$ and $\Gamma = 0.9$.}
     \label{fig_ETD_EE_vs_Bsmax_Bdmax_Joint}
     \vspace{-0.4cm}
     \normalsize 
\end{figure}

\begin{figure}\centering
     \begin{subfigure}[b]{0.35\textwidth}
         \centering
         \includegraphics[width=\textwidth]{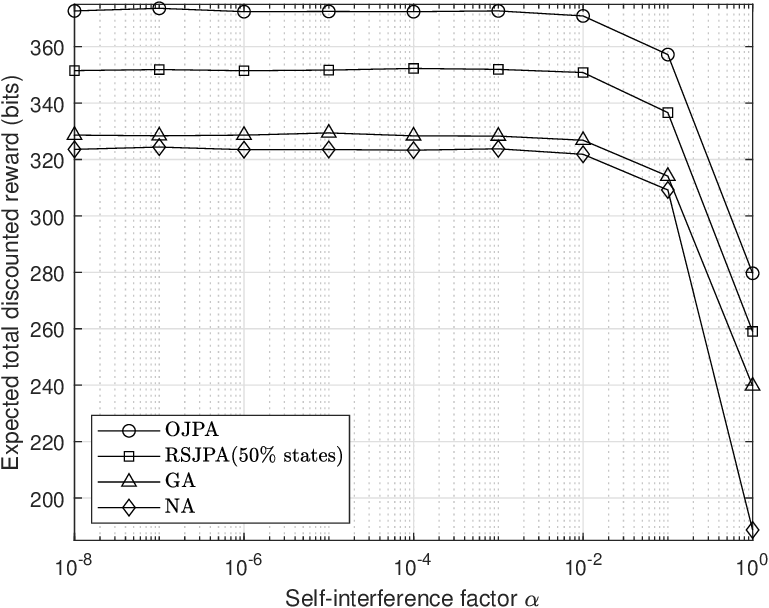}
         \caption{Expected total discounted reward versus $\alpha$ for joint power allocation algorithms where $B_{\textrm{S}}^{\textrm{max}} = B_{\textrm{D}}^{\textrm{max}} = 5$ energy units.}
         \label{fig_ETD_vs_self_interference}
         \vspace{0.3cm}
     \end{subfigure}
       \hfill
     \begin{subfigure}[b]{0.35\textwidth}
         \centering
         \includegraphics[width=\textwidth]{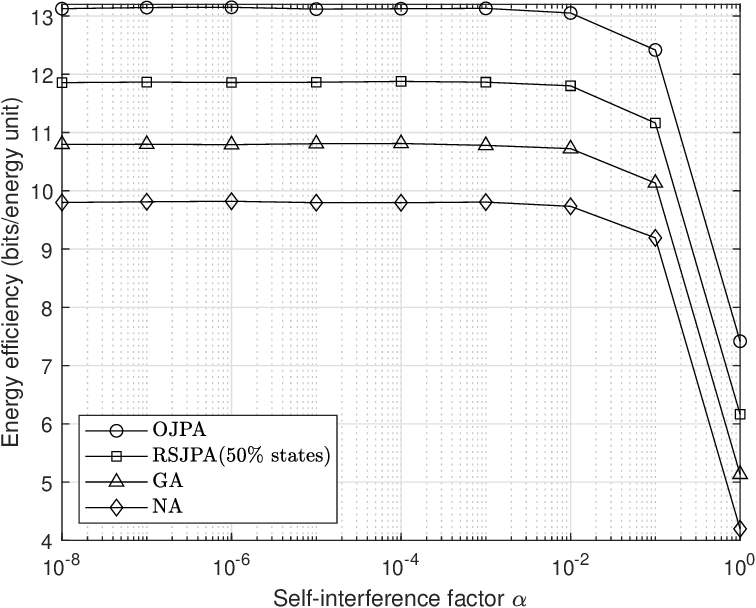}
         \caption{Energy efficiency versus $\alpha$ for joint power allocation algorithms where $B_{\textrm{S}}^{\textrm{max}} = B_{\textrm{D}}^{\textrm{max}} = 5$ energy units.}
         \label{fig_EE_vs_self_interference}
     \end{subfigure}
   
     \caption{Expected total discounted reward and energy efficiency versus self-interference.}
     \label{fig_ETD_EE_vs_self_interference}
     \vspace{-0.4cm}
\end{figure}

Figs. \ref{fig_ETD_vs_Bsmax} and \ref{fig_ETD_vs_Bdmax} plot expected total discounted reward versus $B_{\textrm{S}}^{\textrm{max}}$ and $B_{\textrm{D}}^{\textrm{max}}$, respectively and compares the performance of algorithms OJPA, RSJPA, GA, and NA.
The corresponding energy efficiency plots are shown in Figs. \ref{fig_EE_vs_Bsmax} and \ref{fig_EE_vs_Bdmax}, respectively. We observe that as $B_{\textrm{S}}^{\textrm{max}}$ or $B_{\textrm{D}}^{\textrm{max}}$ increases, the expected total discounted reward improves.  
When $B_{\textrm{S}}^{\textrm{max}}$ and $B_{\textrm{D}}^{\textrm{max}}$ is larger, nodes $\textrm{S}$ and $\textrm{D}$ can store more harvested energy which leads to more expected total transmitted secure bits.
If we consider energy efficiency in  Figs. \ref{fig_EE_vs_Bsmax} and \ref{fig_EE_vs_Bdmax}, energy efficiency also improves when  $B_{\textrm{S}}^{\textrm{max}}$ increases, however, the same observation is not true when $B_{\textrm{D}}^{\textrm{max}}$ increases. When  $B_{\textrm{D}}^{\textrm{max}}$ increases, energy efficiency first improves, however, later degrades as $B_{\textrm{D}}^{\textrm{max}}$ increases further.  As  $B_{\textrm{D}}^{\textrm{max}}$ increases, the possibility of allocating destination jamming power increases which initially leads to better energy efficiency due to increased expected total transmitted secure bits, however, if jamming power further increases, energy efficiency decreases due to increased total power expenditure.

Fig. \ref{fig_ETD_vs_self_interference} shows the expected total discounted reward versus $\alpha$ for the algorithms OJPA, RSJPA, GA, and NA.
The corresponding energy efficiency is plotted in Fig \ref{fig_EE_vs_self_interference}. Both  Figs \ref{fig_ETD_vs_self_interference} and \ref{fig_EE_vs_self_interference} shows the similar trend with $\alpha$. When $\alpha$ is sufficiently small, we do not see any impact of $\alpha$ on the expected total discounted reward or energy efficiency because at very small $\alpha$, self-interference is negligible as compared to the AWGN at node $\textrm{D}$. However, when  $\alpha$ is large enough, the performance degrades as the influence of self-interference on node $\textrm{D}$ becomes comparable to that of the AWGN.

\begin{figure}
 \centering 
 \includegraphics[width = 2.5in]{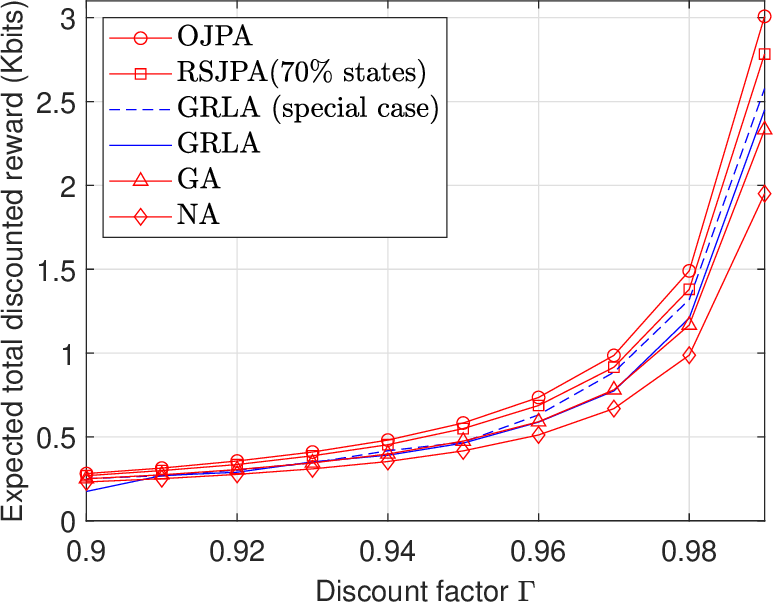} 
 \caption{Expected total discounted reward versus $\Gamma$ when $p = q = 0.5$.}
 \label{fig_ETD_vs_Gamma_GA}
 \vspace{-0.4cm}
\end{figure}

Fig. \ref{fig_ETD_vs_Gamma_GA} compares the performance of proposed algorithms (OJPA and RSJPA) with that of a genetic reinforcement learning algorithm (GRLA), GA, and NA. 
The GRLA combines the standard genetic algorithm \cite{zhao2023_genetic} and with a RL approach, similar to that described in \cite{moriarty1999evolutionary}. In GRLA, the initial population is randomly initialized, and the fitness function of each chromosome is evaluated by summing the value functions for each state corresponding to the actions in that chromosome. The value function for each state is determined iteratively using the Bellman equation (\ref{eq_bellman}) until convergence. 
The rest of the GRLA then applies standard genetic operations, including selection, crossover, and mutation with a population size of $10^3$, a maximum iteration of $10^3$, a mutation probability of $0.01$, a crossover probability of $0.6$, and the roulette wheel selection as the selection strategy. Additionally, Fig. \ref{fig_ETD_vs_Gamma_GA} shows the performance of GRLA in its special case, where the initial population includes a chromosome derived from the GA. This approach enhances the performance of GRLA. We observe that both GRLA and GRLA (special case) outperform the GA and NA because they take into account the cumulative reward when calculating the fitness function. However, their performance is still inferior to the proposed OJPA and RSJPA algorithms when considering $70\%$ of the states.

\color{black}

\begin{figure}
 \centering 
 \includegraphics[width = 2.5in]{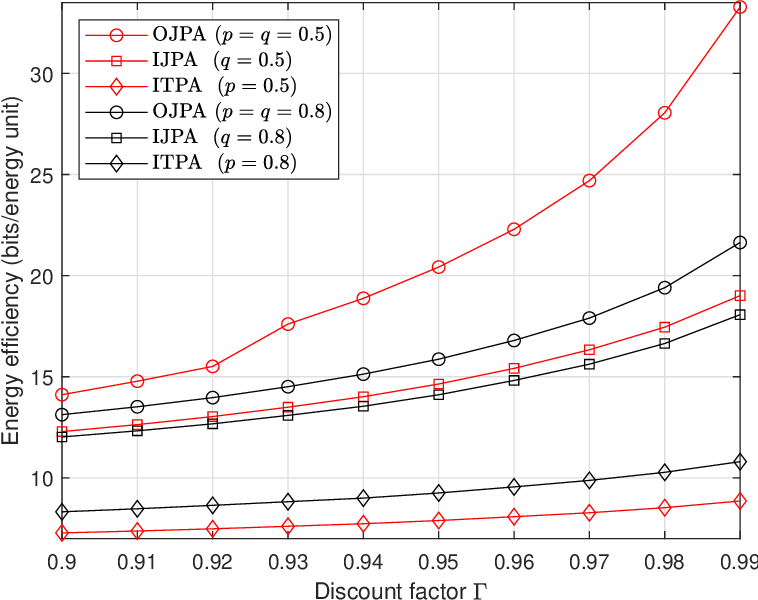} 
 \caption{Energy efficiency versus $\Gamma$ for OJPA algorithm and two IPA algorithms IJPA and ITPA for different $p$ and $q$ with $B_{\textrm{S}}^{\textrm{max}} = B_{\textrm{D}}^{\textrm{max}} = 5$ energy units.}
 \label{fig_EE_vs_Gamma_allcases}
 \vspace{-0.4cm}
\end{figure}

Fig. \ref{fig_EE_vs_Gamma_allcases} plots the energy efficiency versus $\Gamma$ and compares OJPA algorithm with that of the two IPA algorithms (ITPA and IJPA). 
It reveals that for any combination of discount factor and EH probability, the OJPA algorithm performs the best. This means that jointly allocating power optimally for node $\textrm{S}$ and node $\textrm{D}$ is more energy efficient than allocating power optimally either for node $\textrm{S}$ or node $\textrm{D}$ while one of these nodes transmitting power at a fixed rate. 
The OJPA algorithm can utilize energies from the batteries of both node $\textrm{S}$ and node $\textrm{D}$ optimally, whereas, in the IPA algorithms, the energy expenditure from one of the batteries is always fixed at the same value. As the OJPA algorithm is more flexible in utilizing energy from both batteries, the energy expenditure is less as compared to the IPA algorithms while maximizing the expected total discounted reward, hence, the performance of the OJPA algorithm is better than the performance of the IPA algorithms (IJPA and ITPA).

Between the IJPA and ITPA algorithms, the IJPA algorithm is more energy efficient than the ITPA algorithm.
The ITPA algorithm optimizes transmit power while keeping jamming power fixed. Whereas, the IJPA algorithm optimizes jamming power with a fixed transmit power. As we maximize the expected total transmitted secure bits, the ITPA algorithm leads to more energy consumption due to fixed jamming power since fixing a particular jamming power only decreases the eavesdropping rate not the useful data rate. In contrast, by utilizing a fixed transmit power and allocating optimal energy for jamming in the IJPA algorithm, the useful data transmission improves, leading to improved energy efficiency for the IJPA algorithm. 
We also find that the energy efficiency decreases with the increasing EH probability for the OJPA and IJPA algorithms. This observation is similar to that in Fig. \ref{fig_ETD_EE_vs_EH_Joint_varypq} for the OJPA algorithm. In the case of the ITPA algorithm, the observation is just the opposite of the OJPA and IJPA algorithms. Though the IJPA algorithm is more energy efficient than the ITPA algorithm, the ITPA algorithm can take better advantage of increased EH probability than the IJPA algorithm as the energy efficiency of the ITPA algorithm improves with the increasing EH probability.

\begin{figure}
 \centering 
 \includegraphics[width = 2.5in]{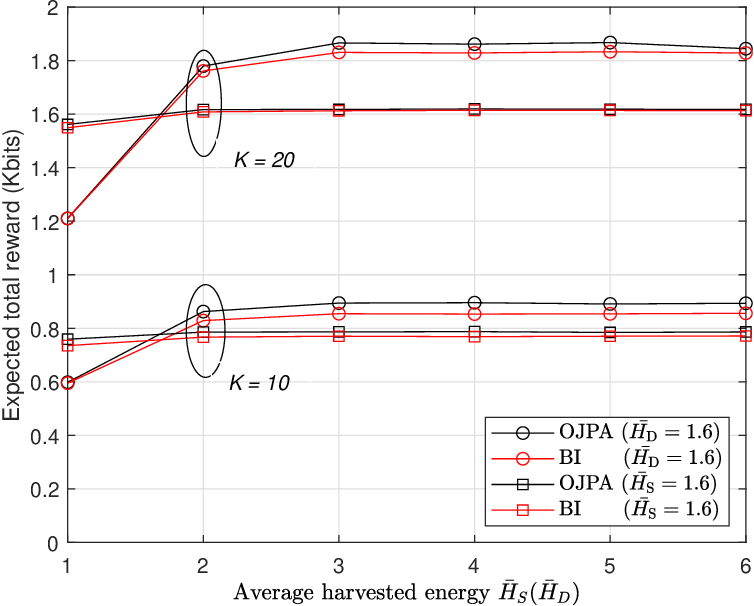} 
 \caption{Expected total reward versus average harvested energy $\bar{H}_\textrm{S} (\bar{H}_\textrm{D})$ for OJPA and BI algorithms with $B_{\textrm{S}}^{\textrm{max}} = B_{\textrm{D}}^{\textrm{max}} = 5$ energy units.}
 \label{fig_ETD_vs_AvgEH_PIvsBI}
 \vspace{-0.4cm}
\end{figure}

To check how optimal policy changes for various timelines, we compare the performance of our system, where the network lifetime $K$ is a random variable, with that of the same system where the network lifetime is a finite constant. In the former case, the joint transmit and jamming power optimization is formulated as an infinite-horizon MDP problem, whereas in the latter case, the joint transmit and jamming power optimization is formulated as a finite-horizon MDP problem. In a finite-horizon problem, the optimal policy is obtained using the Backward Induction (BI)  algorithm \cite{sutton1998RL, Octavia2019 }. For a fair comparison, we assume the mean of the network
lifetime in the OJPA algorithm is the same as the network
lifetime in the BI algorithm. In Fig. \ref{fig_ETD_vs_AvgEH_PIvsBI}, we examine the expected total reward for both the algorithms versus average harvested energy $\bar{H}_\textrm{S}$ and $\bar{H}_\textrm{D}$ individually at $\textrm{S}$ and $\textrm{D}$, respectively. Average harvested energy at $\textrm{S}$ and $\textrm{D}$ are defined as $\bar{H}_\textrm{S} = p E_\textrm{S}$ and $\bar{H}_\textrm{D} = q E_\textrm{D}$. Fig. \ref{fig_ETD_vs_AvgEH_PIvsBI} plots two cases in the same figure: i) $\bar{H}_\textrm{S}$ is on the $x$-axis with $p=0.5$ when $\bar{H}_\textrm{D} = 1.6$ $(q = 0.8, E_\textrm{D}=2)$ and ii) $\bar{H}_\textrm{D}$ is on the $x$-axis with $q=0.5$ when $\bar{H}_\textrm{S} = 1.6$ $(p = 0.8, E_\textrm{S} = 2)$. 
We note that in the $y$-axis, we plot the expected total reward instead of the expected total discounted reward as the network lifetime is known. The plots are also shown in two network lifetime scenarios when $K=10$ and $K=20$.

We observe from Fig. \ref{fig_ETD_vs_AvgEH_PIvsBI} that the OJPA algorithm performs better than the BI algorithm. This is because the BI algorithm does not take into account the randomness of the network lifetime which OJPA does. We also notice that the expected total reward saturates after the average harvested energy reaches a certain threshold. The saturation occurs when the harvested energy exceeds the battery capacity of either $\textrm{S}$ or $\textrm{D}$, and the surplus harvested energy is lost due to battery overflow.
Similar to Fig. \ref{fig_ETD_vs_Gamma_P&Q}, we observe that the increase in harvested energy at $\textrm{S}$ is more beneficial. We also notice that the expected total reward increases as the network lifetime increases from $K=10$ to  $K=20$, however, the nature of the graphs remains the same.

For the quantitative analysis of the computational complexities of the OJPA and RSJPA algorithms, we measured the execution time of the planning phase (RL training phase) of these algorithms in a typical desktop computer. It was found that the average computational time for the planning phase of OJPA and RSJPA algorithms is 25.04 seconds and 6.09 seconds, respectively, for the system parameter considered in Fig. \ref{fig_ETD_vs_Gamma}. This suggests that the RSJPA algorithm can reduce computational time by 75.67 percent as compared to the OJPA algorithm.

The requirement of computation power to train the proposed RL models is a critical consideration for their suitability in distributed networks with low-complexity nodes. Computation energy can affect the energy efficiency of the algorithms which is not currently considered in (\ref{EE}). 
Thus, in our future work, we will measure the energy efficiency of the proposed algorithms considering computation energy, providing a better trade-off between the secrecy performance and the energy efficiency of the optimal and sub-optimal algorithms.

\section{Conclusion}
In this paper, we consider a wireless network with a source and a destination in the presence of an eavesdropper where both the source and the destination are equipped with EH devices with limited battery and the destination has full-duplex jamming capability. 
We study the problem of joint transmit and jamming power allocation at the source and the destination, respectively, to maximize the long-term expected total transmitted secure bits until the network stops functioning. We formulate infinite-horizon Markov decision process problems for the joint optimization solutions. An optimal algorithm OJPA and a sub-optimal algorithm RSJPA are proposed using the PI algorithm in the RL framework.  The results are compared with other sub-optimal algorithms, i.e., GA and NA. 
Computational complexities of the joint power allocation algorithms are provided. We observe that the proposed RSJPA algorithm achieves nearly optimal secrecy performance with significantly less computational complexity than the OJPA algorithm as it adopts a hybrid approach between the OJPA algorithm and the GA. When we compare the energy efficiency of the OJPA algorithm with two individual power allocation algorithms ITPA and IJPA, and all the sub-optimal joint power allocation algorithms RSJPA, GA, and NA, the OJPA performs the best. Hence, the OJPA algorithm not only provides the best secrecy performance, but also the most energy efficient. 
The secrecy performance of the OJPA algorithm is also compared with the GRLA and BI algorithms, where OJPA achieves the best performance.
Furthermore, the RSJPA algorithm, though sub-optimal, can be a balanced choice between the computational complexity and secrecy performance for joint power allocation. It is observed that the RSJPA algorithm with considering only $50$ percent of total number of states can reduce computational time by around $75$ percent as compared to the OJPA algorithm.


\bibliographystyle{IEEEtran}
\bibliography{IEEEabrv,PA_RL}

\begin{thebibliography}{10}
\providecommand{\url}[1]{#1}
\csname url@samestyle\endcsname
\providecommand{\newblock}{\relax}
\providecommand{\bibinfo}[2]{#2}
\providecommand{\BIBentrySTDinterwordspacing}{\spaceskip=0pt\relax}
\providecommand{\BIBentryALTinterwordstretchfactor}{4}
\providecommand{\BIBentryALTinterwordspacing}{\spaceskip=\fontdimen2\font plus
\BIBentryALTinterwordstretchfactor\fontdimen3\font minus \fontdimen4\font\relax}
\providecommand{\BIBforeignlanguage}[2]{{%
\expandafter\ifx\csname l@#1\endcsname\relax
\typeout{** WARNING: IEEEtran.bst: No hyphenation pattern has been}%
\typeout{** loaded for the language `#1'. Using the pattern for}%
\typeout{** the default language instead.}%
\else
\language=\csname l@#1\endcsname
\fi
#2}}
\providecommand{\BIBdecl}{\relax}
\BIBdecl

\bibitem{xu2015_WSN_intro}
W.~Xu, Y.~Zhang, Q.~Shi, and X.~Wang, ``Energy management and cross layer optimization for wireless sensor network powered by heterogeneous energy sources,'' \emph{IEEE Trans. Wireless Commun.}, vol.~14, no.~5, pp. 2814--2826, May 2015.

\bibitem{sah2022_WSN_intro}
D.~K. Sah, A.~Hazra, R.~Kumar, and T.~Amgoth, ``Harvested energy prediction technique for solar-powered wireless sensor networks,'' \emph{IEEE Sensors J.}, vol.~23, no.~8, pp. 8932--8940, Apr. 2022.

\bibitem{sudevalayam2010EH_paper}
S.~Sudevalayam and P.~Kulkarni, ``Energy harvesting sensor nodes: Survey and implications,'' \emph{IEEE Commun. Surveys Tuts.}, vol.~13, no.~3, pp. 443--461, 3rd Quart., 2011.

\bibitem{ku2015EH_paper}
M.-L. Ku, W.~Li, Y.~Chen, and K.~J. Ray~Liu, ``Advances in energy harvesting communications: Past, present, and future challenges,'' \emph{IEEE Commun. Surveys Tuts.}, vol.~18, no.~2, pp. 1384--1412, 2nd Quart., 2016.

\bibitem{dohler2013learning}
P.~Blasco, D.~Gunduz, and M.~Dohler, ``A learning theoretic approach to energy harvesting communication system optimization,'' \emph{IEEE Trans. Wireless Commun.}, vol.~12, no.~4, pp. 1872--1882, Apr. 2013.

\bibitem{azarhava2020_conv_optimization}
H.~Azarhava and J.~M. Niya, ``Energy efficient resource allocation in wireless energy harvesting sensor networks,'' \emph{IEEE Wireless Commun. Lett.}, vol.~9, no.~7, pp. 1000--1003, Jul. 2020.

\bibitem{salari2023_conv_optimization}
S.~Salari and F.~Chan, ``Maximizing the sum-rate of secondary cognitive radio networks by jointly optimizing beamforming and energy harvesting time,'' \emph{IEEE Trans. Veh. Technol.}, vol.~72, no.~6, pp. 8128--8133, Jun. 2023.

\bibitem{Octavia2019}
R.~Wang, A.~Yadav, E.~A. Makled, O.~A. Dobre, R.~Zhao, and P.~K. Varshney, ``Optimal power allocation for full-duplex underwater relay networks with energy harvesting: A reinforcement learning approach,'' \emph{IEEE Wireless Commun. Lett.}, vol.~9, no.~2, pp. 223--227, Feb. 2020.

\bibitem{yadav2017_TCOM_intro}
A.~Yadav, M.~Goonewardena, W.~Ajib, O.~A. Dobre, and H.~Elbiaze, ``Energy management for energy harvesting wireless sensors with adaptive retransmission,'' \emph{IEEE Trans. Commun.}, vol.~65, no.~12, pp. 5487--5498, Dec. 2017.

\bibitem{Ajib_VTC}
A.~Hentati, F.~Abdelkefi, and W.~Ajib, ``Energy allocation for sensing and transmission in wsns with energy harvesting {T}x/{R}x,'' in \emph{Proc. IEEE Veh. Technol. Conf.}, Boston, MA, USA, Sep. 2015, pp. 1--5.

\bibitem{sutton1998RL}
R.~S. Sutton and A.~G. Barto, \emph{Reinforcement Learning: An Introduction}.\hskip 1em plus 0.5em minus 0.4em\relax Cambridge, MA, USA: MIT Press, 2018.

\bibitem{puterman2014markov}
M.~L. Puterman, \emph{Markov Decision Processes: Discrete Stochastic Dynamic Programming}.\hskip 1em plus 0.5em minus 0.4em\relax John Wiley \& Sons, 2014.

\bibitem{wong2012ICC}
S.~Mao, M.~H. Cheung, and V.~W.~S. Wong, ``An optimal energy allocation algorithm for energy harvesting wireless sensor networks,'' in \emph{Proc. IEEE Int. Conf. Commun.}, Ottawa, ON, Canada, Jun. 2012, pp. 265--270.

\bibitem{kashef2012optimal}
M.~Kashef and A.~Ephremides, ``Optimal packet scheduling for energy harvesting sources on time varying wireless channels,'' \emph{J. Commun. Netw.}, vol.~14, no.~2, pp. 121--129, Apr. 2012.

\bibitem{wong2014joint}
S.~Mao, M.~H. Cheung, and V.~W.~S. Wong, ``Joint energy allocation for sensing and transmission in rechargeable wireless sensor networks,'' \emph{IEEE Trans. Veh. Technol.}, vol.~63, no.~6, pp. 2862--2875, Jul. 2014.

\bibitem{PI_heydari2016}
A.~Heydari, ``Analyzing policy iteration in optimal control,'' in \emph{Proc. Amer. Control Conf.}, Boston, MA, USA, Jul. 2016, pp. 5728--5733.

\bibitem{zou2016survey}
Y.~Zou, J.~Zhu, X.~Wang, and L.~Hanzo, ``A survey on wireless security: Technical challenges, recent advances, and future trends,'' \emph{Proc. IEEE}, vol. 104, no.~9, pp. 1727--1765, Sep. 2016.

\bibitem{zhou2012PLS}
X.~Zhou, B.~Maham, and A.~Hjorungnes, ``Pilot contamination for active eavesdropping,'' \emph{IEEE Trans. Wireless Commun.}, vol.~11, no.~3, pp. 903--907, Mar. 2012.

\bibitem{shiu2011PLS}
Y.-S. Shiu, S.~Y. Chang, H.-C. Wu, S.~C.-H. Huang, and H.-H. Chen, ``Physical layer security in wireless networks: A tutorial,'' \emph{IEEE Wireless Commun.}, vol.~18, no.~2, pp. 66--74, Apr. 2011.

\bibitem{wyner_wiretap}
A.~D. Wyner, ``The wire-tap channel,'' \emph{Bell Syst. Tech. J}, vol.~54, no.~8, pp. 1355--1387, Oct. 1975.

\bibitem{hoseini2023GC}
S.~A. Hoseini, F.~Bouhafs, N.~Aboutorab, P.~Sadeghi, and F.~den Hartog, ``Cooperative jamming for physical layer security enhancement using deep reinforcement learning,'' in \emph{Proc. IEEE Glob. Commun. Conf. Workshops}, Kuala Lumpur, Malaysia, Dec. 2023, pp. 1838--1843.

\bibitem{yang2020_PLS_drl_TWC}
H.~Yang, Z.~Xiong, J.~Zhao, D.~Niyato, L.~Xiao, and Q.~Wu, ``Deep reinforcement learning-based intelligent reflecting surface for secure wireless communications,'' \emph{IEEE Trans. Wireless Commun.}, vol.~20, no.~1, pp. 375--388, Jan. 2021.

\bibitem{saleem2022_PLS_drl}
R.~Saleem, W.~Ni, M.~Ikram, and A.~Jamalipour, ``Deep-reinforcement-learning-driven secrecy design for intelligent-reflecting-surface-based {6G-IoT} networks,'' \emph{IEEE Internet Things J.}, vol.~10, no.~10, pp. 8812--8824, May 2023.

\bibitem{liu2024secrecy_DRL}
S.~Liu, X.~Liu, X.~Du, and M.~Guizani, ``Smart jamming for secrecy: Deep reinforcement learning enabled secure visible light communication,'' \emph{IEEE Trans. Wireless Commun.}, vol.~23, no.~12, pp. 17\,915--17\,928, Dec. 2024.

\bibitem{li2022DRL}
B.~Li, T.~Shi, W.~Zhao, and N.~Wang, ``Reinforcement learning-based intelligent reflecting surface assisted communications against smart attackers,'' \emph{IEEE Trans. Commun.}, vol.~70, no.~7, pp. 4771--4779, Jul. 2022.

\bibitem{yang2024EH_DRL}
H.~Yang, K.~Lin, L.~Xiao, Y.~Zhao, Z.~Xiong, and Z.~Han, ``Energy harvesting uav-ris-assisted maritime communications based on deep reinforcement learning against jamming,'' \emph{IEEE Trans. Wireless Commun.}, vol.~23, no.~8, pp. 9854--9868, Aug. 2024.

\bibitem{qian2022secrecy}
L.~P. Qian, W.~Zhang, H.~Zhang, Y.~Wu, and X.~Yang, ``Secrecy capacity maximization for uav aided noma communication networks,'' in \emph{Proc. IEEE Int. Conf. Commun.}, Seoul, Korea, May. 2022, pp. 3130--3135.

\bibitem{insoo2019FD}
Q.~V. Do, T.-N.-K. Hoan, and I.~Koo, ``Optimal power allocation for energy-efficient data transmission against full-duplex active eavesdroppers in wireless sensor networks,'' \emph{IEEE Sensors J.}, vol.~19, no.~13, pp. 5333--5346, Jul. 2019.

\bibitem{octavia2020VTC}
R.~Wang, E.~A. Makled, A.~Yadav, O.~A. Dobre, and R.~Zhao, ``Reinforcement learning-based energy-efficient power allocation for underwater full-duplex relay network with energy harvesting,'' in \emph{Proc. IEEE Veh. Technol. Conf.}, Victoria, BC, Canada, 18 Nov. - 16 Dec. 2020, pp. 1--5.

\bibitem{bellman1957markovian}
R.~Bellman, ``{A Markovian decision process},'' \emph{J. Math. Mechanics}, vol.~6, no.~5, pp. 679--684, 1957.

\bibitem{ahmed2012power}
I.~Ahmed, A.~Ikhlef, R.~Schober, and R.~K. Mallik, ``Power allocation in energy harvesting relay systems,'' in \emph{Proc. IEEE Veh. Technol. Conf.}, Yokohama, Japan, May 2012, pp. 1--5.

\bibitem{ahmed2013joint}
I.~Ahmed, A.~Ikhlef, R.~Schober, and R.~K. Mallik, ``Joint power allocation and relay selection in energy harvesting {AF} relay systems,'' \emph{IEEE Wireless Commun. Lett.}, vol.~2, no.~2, pp. 239--242, Apr. 2013.

\bibitem{poor2009secrecy}
L.~Dong, Z.~Han, A.~P. Petropulu, and H.~V. Poor, ``Improving wireless physical layer security via cooperating relays,'' \emph{IEEE Trans. Signal Process}, vol.~58, no.~3, pp. 1875--1888, Mar. 2009.

\bibitem{poor2008secure}
Y.~Liang, H.~V. Poor, and S.~Shamai, ``Secure communication over fading channels,'' \emph{IEEE Trans. Inf. Theory}, vol.~54, no.~6, pp. 2470--2492, Jun. 2008.

\bibitem{poor2014security}
L.~Wang, K.~J. Kim, T.~Q. Duong, M.~Elkashlan, and H.~V. Poor, ``Security enhancement of cooperative single carrier systems,'' \emph{IEEE Trans. Inf. Forensics Security}, vol.~10, no.~1, pp. 90--103, Jan. 2015.

\bibitem{kundu2021ergodic}
C.~Kundu and M.~F. Flanagan, ``Ergodic secrecy rate of optimal source selection in a multi-source system with unreliable backhaul,'' \emph{IEEE Wireless Commun. Lett.}, vol.~10, no.~5, pp. 1118--1122, Feb. 2021.

\bibitem{shashi2024TVT}
S.~B. Kotwal, C.~Kundu, S.~Modem, and M.~F. Flanagan, ``Transmitter selection for secrecy in frequency-selective fading with multiple eavesdroppers and wireless backhaul links,'' \emph{IEEE Trans. Veh. Technol.}, vol.~73, no.~1, pp. 860--875, Jan. 2024.

\bibitem{PIcomplexity}
Y.~Mansour and S.~Singh, ``On the complexity of policy iteration,'' in \emph{Proc. Int. Conf. Uncertainty Artif. Intell.}, 1999, pp. 401--408.

\bibitem{ieee1997wireless}
\emph{IEEE Standard for Local and Metropolitan Area Networks--Part 15.4: Low-Rate Wireless Personal Area Networks (LR-WPANs)}, IEEE Standard 802.15.4-2011, 2011.

\bibitem{zhao2023_genetic}
Y.~Zhao, Y.~Zhu, and S.~Wang, ``User scheduling in wireless networks for deterministic service: An efficient genetic algorithm method,'' \emph{IEEE Netw. Lett.}, vol.~6, no.~1, pp. 1--5, Dec. 2023.

\bibitem{moriarty1999evolutionary}
D.~E. Moriarty, A.~C. Schultz, and J.~J. Grefenstette, ``Evolutionary algorithms for reinforcement learning,'' \emph{J. Artif. Intell. Res.}, vol.~11, pp. 241--276, Sep. 1999.

\end{thebibliography}

\end{document}